\documentclass[11pt,twoside]{book} 
\usepackage{asp2008n}
\usepackage{times}
\usepackage{lscape}
\usepackage{epsf}

\usepackage{graphicx}
\usepackage{amssymb}
\usepackage{longtable}
\usepackage[figuresright]{rotating}
\usepackage{multirow}

\newcommand{\simless}{\mathbin{\lower 3pt\hbox {$\rlap{\raise 5pt\hbox{$\char'074$}}\mathchar"7218$}}}

\mainmatter

\newlength{\deftabcolsep}
\begin{document}

\title{Orion Outlying Clouds}   %%% Fill in title
\author{Juan M. Alcal{\' a}, Elvira Covino and Silvio Leccia}   %%% Fill in author names
\affil{INAF- Osservatorio Astronomico di Capodimonte,\\
Via Moiariello 16, 80131, Naples, Italy}    %%% Fill in author affiliations

\begin{abstract} %%% Abstract to run on from here.
In this chapter we review the properties of the Orion outlying clouds at
$b < -21^\circ$. These clouds are located far off the Orion giant molecular
cloud complex and are in most cases small cometary-shaped clouds, with their
head pointing back towards the main Orion clouds. A wealth of data indicate
that star formation is ongoing in many of these clouds. The star formation
in these regions might have been triggered due to the strong impact of the
massive stars in the Orion OB association. Some of the clouds discussed
here may be part of the Orion-Eridanus bubble. An overview on each individual
cloud is given. A synthesis of the Pre-Main Sequence stars discovered in
these clouds is presented. We also discuss the millimeter and centimeter
data and present a review of the outflows and Herbig-Haro objects so far
discovered in these clouds.

\end{abstract}

%%% MAIN BODY OF TEXT GOES HERE. CONSULT "INSTRUCTIONS FOR AUTHORS USING
%%% LATEX2E MARKUP", SECTIONS 2.3-2.6 FOR HELP WITH EQUATIONS, FIGURES,
%%% AND TABLES.

%\section{}   %%% Top level section head (remove "%" symbol)
%\subsection{}   %%% Second level section head (remove "%" symbol)
%\subsubsection{}   %%% Lowest level section head (remove "%" symbol)
%\section*{}	%%% Unnumbered top level section head (remove "%" symbol)
%\subsection*{}   %%% Unnumbered second level section head (remove "%" symbol)

% ----------------------------------------------------------------------------------
% ----------------------------------------------------------------------------------

\section{Introduction}
\label{intro}

The Orion Outlying Clouds  are typically small clouds located in the
outskirts of the Orion molecular cloud complex. Most of the known clouds
of this kind are located to the west of the Orion OB association.
They often appear as small cometary clouds, which point back towards the
Orion OB association
and in which star formation might have been triggered due to the strong impact
of the massive stars in the Orion OB association. The illumination of dense
clumps in molecular clouds by OB stars could be responsible for their collapse
and subsequent star formation.
The UV radiation from the OB stars may sweep the molecular cloud material
into a cometary shape with a dense core located at the head
of the cometary clouds. Bright Rimmed Clouds (BRCs) associated with HII
regions are examples where the UV flux of a nearby OB star ionizes the
external layers of the cloud and causes the BRC to collapse.
The shape of a BRC (curved or cometary-like) is governed by the ionization
fronts from the OB stars that, compressing the head of the BRC, enhance the
density of the outer layers of the cloud. This process
may lead to sequential star formation.
\citet{Sugi91} and \citet{Sugi94} catalogue 89 BRCs associated with IRAS point
sources in both the northern and southern hemispheres, some of which are
also associated with Herbig-Haro objects and molecular outflows.
These BRCs are sites in which triggered star formation might have
taken place.

%%%%%%%%%%%%%%%%%% Fig 1 %%%%%%%%%%%%%%%%%%%%
\begin{figure}[!ht]
% \plotfiddle{clouds_new.ps}{9.3cm}{0}{75}{75}{-260}{-165}
% \plotfiddle{fig1.ps}{9.3cm}{0}{75}{75}{-260}{-165}
\begin{center}
\includegraphics[width=\textwidth]{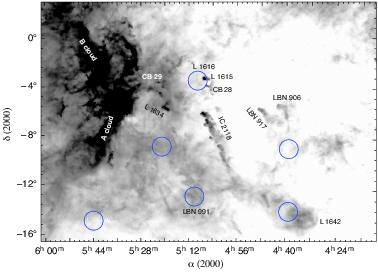}
\caption{Map of the 100~$\mu$m IRAS dust emission of the Orion OB association.
The main Orion molecular clouds A and B are indicated. The Orion outlying
clouds discussed in this chapter are labelled. The big circles represent
the  ROSAT X-ray clumps that satisfy the \citet{Sterz95} criteria.}
% \special{psfile=ds9c.ps vscale=70 hscale=70 angle=0 voffset=10 hoffset=-20}
\label{clouds}
\end{center}
\end{figure}
%%%%%%%%%%%%%%%%%%%%%%%%%%%%%%%%%%%%%%%%%%%%%%%

\begin{table}[!ht]
\vspace{8mm}
\caption{\label{tab:coordinates} Orion outlying clouds discussed in
this chapter}
\label{coordinates}
% \smallskip
\begin{center}
{\small
\begin{tabular}{lccccll}
\tableline
\noalign{\smallskip}
Cloud & $\alpha$(2000)   & $\delta$(2000)  & $l$      & $b$      &  other IDs  & refs.  \\
      & h~~~m~~~s   & $^{\circ}$~~~~'~~~~''& ($^\circ$) & ($^\circ$) &         &        \\
\noalign{\smallskip}
\tableline
\noalign{\smallskip}
CB~29               & 05:22:12 & --03:41:34  & 205.8 & --21.5 &                  & 1     \\
L~1634              & 05:19:49 & --05:52:05  & 207.6 & --23.0 & MBM~110          & 2, 3  \\
L\,1616$^\dagger$    & 05:07:00 & --03:21:06  & 203.5 & --24.7 &                  & 2     \\
L\,1615$^\dagger$    & 05:05:30 & --03:26:00  & 203.4 & --25.1 &                  & 2     \\
CB~28               & 05:06:20 & --03:56:22  & 204.0 & --25.1 & LBN~923          & 1, 2  \\
IC~2118             & 05:03:55 & --08:23:59  & 208.1 & --27.7 &                  & 4     \\
LBN~991$^\dagger$   & 05:11:00 & --12:24:00  & 213.1 & --27.8 &                  & 2     \\
LBN~917             & 04:47:42 & --05:55:00  & 203.5 & --30.1 & DIR~203$-$32     & 2, 5  \\
LBN~906             & 04:41:00 & --05:24:00  & 202.1 & --31.4 &                  & 2     \\
L~1642$^\dagger$    & 04:35:03 & --14:13:57  & 210.9 & --36.6 & LBN~981, MBM~20  & 2, 3  \\
\noalign{\smallskip}
\tableline
\noalign{\smallskip}
\multicolumn{7}{l}{\parbox{0.9\textwidth}{\footnotesize
    $^\dagger$ coinciding with an X-ray clump (see Sect.~1).}}\\[2ex]
\multicolumn{7}{l}{\parbox{0.9\textwidth}{\footnotesize
    References:
    1:~\citet{Clem88};
    2:~\citet{Lyn62};
    3:~\citet{MagBliMun85};
    4:~\citet{Drey1908}: {\it Index Catalog};
    5:~\citet{reach98}   }}\\
\end{tabular}
}
\begin{flushleft}
\end{flushleft}

\end{center}
\end{table}

In this chapter we discuss the outlying clouds in Orion at $b < -21^\circ$,
focusing mainly on the four best studied cases, namely L\,1615/L\,1616, L\,1634,
the IC~2118 region and L\,1642. However, few aspects of other high-galactic
latitude clouds are also mentioned. Some of these clouds may be inside the
so called Orion-Eridanus bubble, a cavity of the interstellar medium towards
Orion which is filled with hot ionized gas surrounded by an expanding shell
of neutral hydrogen. For a detailed description of the Orion-Eridanus bubble
we refer the reader to \citet{Brown95}. We use the 100~$\mu$m IRAS map in order
to identify the Orion outlying clouds discussed here and refer to Figure~\ref{clouds}
for their spatial location and to Table~\ref{tab:coordinates} for their approximate
coordinates, in order of decreasing Galactic latitude.

In addition to the Orion outlying clouds we also discuss a few topics of
the four small bright-rimmed clouds and cometary globules No. 27, 35, 40
and 41 by  \citet{Ogu98}, which are located in the vicinity of the O7~V star
$\sigma$~Ori.

The PMS objects in each of the outlying clouds and globules are also discussed.
A number of H$\alpha$-emission stars are found on the clouds, but other are
scattered around them. Their coordinates, designations and other informations
are synthesized and reported in a table (Table~\ref{tab:tts}). For many of
these stars a proper motion determination is provided in the catalogue by
\citet{Ducour05}. The objects in the Orion OB1a association are not discussed
here, but they are treated in other chapters of this book. While some
H$\alpha$-emission stars in the outer regions of Orion were discovered in
the objective-prism survey by \citet{Step86}, many others were revealed in
the Kiso H$\alpha$ survey  \citep{Wir91, Nak95}. Other weak H$\alpha$-emission
stars in these regions were identified as optical counterparts of the
ROSAT All-Sky Survey sources in spectroscopic follow-ups
\citep[e.g.][]{Alc96, Alc00}.

The complete sky coverage of the ROSAT All-Sky Survey  permits an unbiased
analysis of the spatial distribution of X-ray active stars, though it is flux
limited. \citet{Sterz95} established a criterion for selecting young star
candidates from ROSAT All-Sky Survey sources based on X-ray hardness ratios
and the ratio of X-ray to optical flux, and used this to trace their spatial
distribution in a $\sim$ 700~deg$^2$ field around the Orion molecular clouds.
The surface density distribution reproduces the major clusters associated with
the OB sub-group associations (OB1a, OB1b, OB1c, and $\lambda$-Ori), and the
spatial extent of the clusters is consistent with dispersal times between
2 and 10~Myrs, e.g. the respective ages of the stellar components in those
regions. The same analysis revealed several overdensities of X-ray sources with
high probability of being low-mass PMS stars \citep{Sterz95, Walt00}.
Several of these X-ray clumps are located far from the main Orion molecular
clouds and some of them are coincident with or very close in position to
the Orion outlying clouds, in particular L\,1615/L\,1616, LBN~991 and L\,1642.
Some of these clumps are indicated with circles in the 100~$\mu$m
IRAS map in Figure~\ref{clouds}. It is interesting to note that L\,1634 and
IC~2118 were not revealed as X-ray enhancements in the analysis by
\citet{Sterz95} (see Figure~\ref{clouds}). This may be an indication that
the X-ray emitting young stellar population in these regions is less
conspicuous, or that the objects are much younger, and/or more embedded,
than in other clouds revealed as X-ray clumps like L\,1616.

The chapter is structured as follows: an overview of the L\,1634,
L\,1616/L\,1615, IC~2118 and L\,1642 clouds and their young
populations are presented in Sections~2, 3, 4 and 5 respectively,
while other small clouds are discussed in Section~6. The four small
clouds around $\sigma$~Ori are then discussed in Section~7, while a
synthesis of the distances of the outlying clouds is presented in
Section~8.  Finally, a summary is presented in Section~9.\\

\section{L~1634}
\label{L1634}

\begin{figure}[!ht]
% \plotfiddle{L1634_dss_color.eps}{8.0cm}{0}{45}{45}{-200}{-80}
% \plotfiddle{L1634_dss_color_detail.eps}{0cm}{0}{20}{20}{60}{60}
%\plotfiddle{fig2_SCALED.eps}{8.0cm}{0}{90}{90}{-190}{-40}
%\plotfiddle{fig2_detail_SCALED.eps}{0cm}{0}{100}{100}{70}{70}
\begin{center}
\includegraphics[width=\textwidth]{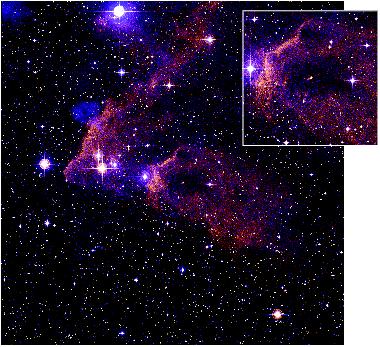}

\caption{The left panel is a color image of the L\,1634 cometary
cloud. The image was produced by the authors by combining three
Digitized Sky Survey (DSS) images. It covers a field of
60$'$$\times$60$'$. North is up and East to the left.
The central region is zoomed in the right panel, which covers
an area of 25$'$$\times$25$'$.}
\label{L1634_color}
\end{center}
\end{figure}

\begin{figure}[!hbt]
% \plotfiddle{L1634.ps}{8cm}{0}{70}{70}{-250}{-160}
%\plotfiddle{fig3.eps}{8cm}{0}{70}{70}{-250}{-160}
\begin{center}
\includegraphics[width=\textwidth]{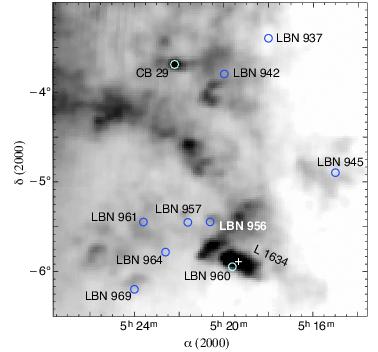}

\caption{Map of the IRAS 100~$\mu$m dust emission of the region
in the L\,1634 cometary cloud. The position of the source
IRAS~05173$-$0555, which drives the spectacular outflow in L\,1634,
is indicated by the white {\it plus} symbol. The position of CB~29
as reported in Table~\ref{tab:coordinates}, as well as of other
Lynds bright nebulae, whose coordinates are listed in
Table~\ref{tab:clouds_L1634}, are indicated with circles.}
\label{clouds_L1634}
\end{center}
\end{figure}

L~1634 \citep{Lyn62} is a small, isolated dark cloud located some
3 degrees to the west of the Orion~A  cloud (cf. Figure~\ref{clouds}).
A color image of the cloud is shown in Figure~\ref{L1634_color}.
The Red Nebulous Object No.~40 \citep[RNO-40 following the nomenclature by][]{Cohen80}
or SFO-16 \citep[following][]{Sugi91} is a small nebulous object in
L~1634. This nebulous object was catalogued as a star with a tailed
nebula by \citet[][their object No. 38]{Gyu77} and coincides with object
PP~27 by \citet{Pars79a}.
The object is seen in the zoomed area in Figure~\ref{L1634_color}.
The L\,1634 cloud also coincides with the MBM~110 cloud \citep{MagBliMun85}
and with the CO emission peak No.~14 reported by \citet{Mad86}.
The Lynds bright nebulae \citep{Lyn65} LBN~956, LBN~957 and LBN~964
are nearby L\,1634, while LBN~960 is located on the cloud
(see Figure~\ref{clouds_L1634}). The coordinates of these bright
nebulae are listed in Table~\ref{tab:clouds_L1634}.

%%%%%%%%%%%%%%%%%% Fig  2 %%%%%%%%%%%%%%%%%%%%
%%%%%%%%%%%%%%%%%%%%%%%%%%%%%%%%%%%%%%%%%%%%%%%

%%%%%%%%%%%%%%%%%% Fig 3 %%%%%%%%%%%%%%%%%%%%
%%%%%%%%%%%%%%%%%%%%%%%%%%%%%%%%%%%%%%%%%%%%%%%

L~1634 was first studied in the CO (J=1 $\rightarrow$ 0 transition)
by \citet{Tor83}. The cloud has been proposed to be either a remnant
of the molecular material from which the nearby Orion OB1 association
formed or a cloud pushed to its present location by the pressure
associated with energetic phenomena accompanying the evolution of
the OB association \citep{Mad86}. L\,1634, which borders Barnard's loop,
is classified as a type-A bright-rimmed cloud rather than as a cometary
cloud, because it has a broad ionization front with no tail \citep{Sugi91}.

The infrared source IRAS~05173$-$0555 is located in this cloud.
\citet{Cohen80} described the object as a purely nebulous, elongated,
very red source typical of Herbig-Haro (HH) objects, with an emission
line spectrum including H$\alpha$, H$\beta$, H$\gamma$,
[N~II], [S~II], [O~I], [N~I], [O~III], and He~I.
Indeed, the L\,1634 cloud is much better known because its core hosts
two spectacular outflows and the Herbig-Haro objects HH~240/241.
The powerful outflow is driven by the aforementioned IRAS source
\citep{Davis97} and most of the research done so far in this
cloud has been focused on the investigation of the outflows.

\subsection{L~1634: PMS Stars}

A number of H$\alpha$ emission-line stars are found on/or very close
to the cloud. Many of these stars are located east of the cloud,
which may be taken as an indication that they are a possible result of
induced star formation.
There are three H$\alpha$ emission-line stars detected in the  survey
by \citet{Step86}, namely StHA~37, StHA~38 and StHA~39.
The star StHA~37 coincides with HBC~83 of the \citet{HeBe88}
catalogue\footnote{In the \citet{HeBe88} catalogue, the entry HBC~83,
corresponding to StHA~37, is erroneously associated with the star V~534~Ori
(which actually coincides with StHA~38).},  which can be also identified
with the IRAS source IRAS~05178$-$0548 and the H$\alpha$-emission star
Kiso~A-0975~52.
StHA~38 coincides with V~534~Ori, while StHA~39 was also detected in X-rays in
the ROSAT All-Sky Survey \citep{Alc96}. Both StHA~38 and StHA~39 can be classified
as T~Tauri stars based on follow-up spectroscopic observations \citep{Dow88, Lee07}.
There are thus at least three on-cloud T~Tauri stars in L\,1634, whose
coordinates are reported in Table~\ref{tab:tts}. Two objects classified as
T~Tauri stars based on follow-ups of the ROSAT All-Sky Survey X-ray
sources \citep{Alc96, Alc00} are scattered in the field of L\,1634 and
CB~29 and are also indicated as T~Tauri stars in Table~\ref{tab:tts}.

%%%%%%%%%%%%%%%%%%%%%%%%%%%%%%%%%%%%%%%%%%% Table %%%%%%%%%%%%%%%%%%%%%%%%%%%%%%%%
\begin{table}[!ht]									     %
\caption{\label{tab:clouds_L1634} Lynds bright nebulae in the field of L\,1634 and CB~29.}    %
\label{coordinates}									     %
% \smallskip				                                         %
\begin{center}  									     %
{\small 										     %
\begin{tabular}{lcccc}  								     %
\tableline										     %
\noalign{\smallskip}									     %
Cloud & RA (2000)   & DEC(2000) 	 & $l$      & $b$	     \\ 		     %
      & h~~~m~~~s   & $^{\circ}$~~~~'~~~~''& ($^\circ$) & ($^\circ$) \\ 		     %
\noalign{\smallskip}									     %
\tableline										     %
\noalign{\smallskip}									     %
LBN~960  & 05:19:37 & $-$05:56:51  & 207.66 & $-$23.11  \\				     %
LBN~956  & 05:20:37 & $-$05:26:51  & 207.29 & $-$22.66  \\				     %
LBN~957  & 05:21:37 & $-$05:27:14  & 207.42 & $-$22.44  \\				     %
LBN~964  & 05:22:37 & $-$05:47:03  & 207.86 & $-$22.37  \\				     %
LBN~961  & 05:23:36 & $-$05:27:04  & 207.66 & $-$22.00  \\				     %
LBN~942  & 05:19:59 & $-$03:47:57  & 205.63 & $-$22.04  \\				     %
LBN~937  & 05:18:00 & $-$03:24:00  & 205.00 & $-$22.29  \\				     %
LBN~945  & 05:15:00 & $-$04:53:51  & 206.06 & $-$23.65  \\				     %
LBN~969  & 05:24:01 & $-$06:12:01  & 208.43 & $-$22.25  \\				     %
\noalign{\smallskip}									     %
\tableline										     %
\end{tabular}										      %
}											       %
\end{center}										      %
\end{table}										      %
%%%%%%%%%%%%%%%%%%%%%%%%%%%%%%%%%%%%%%%%%%%%%%%%%%%%%%%%%%%%%%%%%%%%%%%%%%%%%%%%%

In addition to the previously known PMS stars, we have selected other
H$\alpha$-emission stars within a radius of about 1.3 degrees from the
two cloud centers. This radius was derived considering that 10~Myr old
stars at a distance of 450~pc drifting with a velocity dispersion of
1~km/s spread-out within a radius of about 10~pc from the clouds center.
Using in addition the 2MASS data and the criteria by \citet{Lee05} we
selected 6 H$\alpha$-emission stars that have a high probability of being
T~Tauri stars based on their near-IR colors. Two of these stars
coincide with previously known T~Tauri stars (namely, StHA~37 and StHA~38).
We report the other 4 stars in Table~\ref{tab:tts} as T~Tauri star candidates
in L\,1634.

\subsection{L~1634: Infrared, Centimeter and Millimeter Sources}

The two most interesting IR sources in the L\,1634 cloud are IRAS~05173$-$0555
and LDN~1634~7 \citep{HoLa95,Davis97}. Their coordinates are given in Table~\ref{tab:tts}.
\citet{Reipetal93} measured the source flux at 1300~$\mu$m and confirmed that it is
a deeply embedded source. They also detected cool dust emission toward the IRAS
source. IRAS~05173$-$0555  has a bolometric luminosity of 17~L$_{\odot}$
\citep{Reipetal93, HoLa95} and may be a transition object between the Class-0 and
Class-I phases \citep{Reipetal93,Beltr02}.

\citet{Beltr02} reported observations of the centimeter, millimeter, and sub-millimeter
continuum emission toward the core of L\,1634. They detected five radio continuum
sources at centimeter and millimeter wavelengths. One of these (VLA~3) coincides
with IRAS~05173$-$0555. The other four are listed in Table~\ref{tab:tts} as YSOs
candidates. % in L\,1634.
\citet{Beltr02} found that the sub-millimeter dust emission
around IRAS~05173$-$0555 is resolved and shows two components, a centrally peaked
source plus a considerably extended envelope, while the emission around LDN~1634~7 appears
unresolved. They also found that the dust properties around the two IR sources are
similar. Based on a power-law model of the radial intensity profiles of the
extended emission, they constrain the density distribution around IRAS~05173$-$0555
and find that the best fit is consistent with the predictions of the inside-out
protostellar collapse model. They derive an infall-mass rate of
(2.6--8.0)$\times10^{-5}$ M$_\odot$ yr$^{-1}$  for the free-falling inner region
of the envelope and conclude that the sub-millimeter luminosity, the total
circumstellar mass derived, and the infall-rate estimate are consistent with
IRAS~05173$-$0555 being a Class-0 object.

In a study of bright-rimmed clouds that includes L\,1634, \citet{DeVr02}
reported the results of a millimeter and sub-millimeter molecular line survey.
They found that the appearance of the millimeter CO and HCO$^{+}$ emission is
dominated by the morphology of the shock front in the bright-rimmed clouds.
The HCO$^{+}$ (J = 1--0) emission tends to trace the swept-up gas ridge and
overdense regions, which may be induced to collapse as a result of sequential
star formation.
In the case of L\,1634, the millimeter and sub-millimeter HCO$^{+}$ observations
by \citet{DeVr02} show some evidence of infall asymmetry.

In addition, there are four IRAS sources in the neighborhood of the small
cloud CB~29. These are IRAS~05194$-$0343, IRAS~05194$-$0346, IRAS~05201$-$0341 and
IRAS~05190$-$0348.

\subsection{L~1634: Outflows and Herbig-Haro Objects}

%%%%%%%%%%%%%%%%%% Fig 4 %%%%%%%%%%%%%%%%%%%%
\begin{figure}[p]
% \plotone{L1634h2.eps}
% \plotone{L1634con.eps}
\begin{center}
\includegraphics[width=\textwidth]{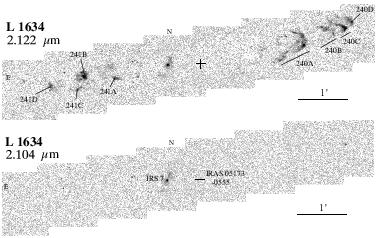}

\caption{Narrow-band images of the L\,1634 outflow at 2.122~$\mu$m (H$_2$ + continuum)
and 2.104~$\mu$m (continuum). The position of the driving source of the outflow
IRAS~05173$-$0555 is marked with a cross. The IR source LDN~1634~7, which might drive
the other outflow in L\,1634, is also indicated as IRS~7.
Adapted from \citet{Davis97}.}
\label{L1634outflow}
\end{center}
\end{figure}
%%%%%%%%%%%%%%%%%%%%%%%%%%%%%%%%%%%%%%%%%%%%%%%

\begin{table}[p]
\caption{\label{tab:hhobjects_L1634} Herbig-Haro Objects in L\,1634$^\ddag$.}
% \label{ic2118clouds}
% \smallskip
 \begin{center}
%{\small
\begin{tabular}{lcccl}
\tableline
\noalign{\smallskip}
Cloud & RA (2000) & DEC(2000)           &  cloud       & driving \\
      & h~~~m~~~s & $^{\circ}$~~~'~~~''  &             &  source  \\
\noalign{\smallskip}
\tableline
\noalign{\smallskip}

HH~240A$^\dagger$ & 05:19:40.45 & $-$05:51:42.3 &  L\,1634 & IRAS~05173$-$0555 \\
HH~240B          & 05:19:38.66 & $-$05:51:23.2 &  L\,1634 & IRAS~05173$-$0555 \\
HH~240C          & 05:19:37.06 & $-$05:51:21.0 &  L\,1634 & IRAS~05173$-$0555 \\
HH~240D          & 05:19:36.56 & $-$05:51:14.0 &  L\,1634 & IRAS~05173$-$0555 \\
HH~241A          & 05:19:57.03 & $-$05:52:24.5 &  L\,1634 & IRAS~05173$-$0555 \\
HH~241B          & 05:19:58.24 & $-$05:52:20.6 &  L\,1634 & IRAS~05173$-$0555 \\
HH~241C          & 05:19:58.83 & $-$05:52:35.6 &  L\,1634 & IRAS~05173$-$0555 \\
HH~241D          & 05:20:01.13 & $-$05:52:33.8 &  L\,1634 & IRAS~05173$-$0555 \\

\noalign{\smallskip}
\tableline
\noalign{\smallskip}

\multicolumn{5}{l}{\parbox{0.8\textwidth}{\footnotesize
$^\ddag$: data adapted from \citet{Davis97, HoLa95}}}\\
\multicolumn{5}{l}{\parbox{0.8\textwidth}{\footnotesize
$^\dagger$: this knot coincides with Kiso~A-0975~48 }}

\end{tabular}

 \end{center}
\end{table}

The core of L\,1634 is probably much better known as the site of the spectacular
bipolar outflow and the Herbig-Haro objects HH 240/241 \citep[in the catalog of][]{Reip00}.
The object was originaly named RNO~40 by \citet{Cohen80}.
HH 240/241   %are Herbig-Haro objects \citep{Cohen80, Bohi93}, which
in the infrared H$_2$
lines are revealed as a powerful bipolar flow \citep{HoLa95, Davis97} extending in the
East-West direction from both sides of the infrared source IRAS~05173$-$0555
(cf. Figure~\ref{L1634outflow}). This source is indicated to be driving the
outflow by the CO (J=3-2) spectra obtained by \citet{Davis97} and has a steeply
rising spectral energy distribution from 12 to 100~$\mu$m \citep{Coh85}.
The L\,1634 powerful outflow has a total length of about 6', corresponding to a
projected length of about 0.8~pc at a distance of 450~pc. The bright HH knots
A, B, C and D identified by \citet{Davis97} along the outflow
are shown in Figure~\ref{L1634outflow}, and their coordinates are reported in
Table~\ref{tab:hhobjects_L1634}. The object catalogued as the emission-line
star Kiso~A-0975~48 coincides with the Herbig-Haro object HH~240A.

\citet{Nisi02} performed an extensive 1-2.5~$\mu$m spectroscopic survey of the bright
HH knots (A, B, C and D) identified by \citet{Davis97} along the two Herbig-Haro chains
HH240-HH241, i.e. the blue-shifted (HH~240) and red-shifted (HH~241) lobes of the
bipolar outflow in L\,1634, respectively (see Figure~\ref{L1634outflow}). They found
that the spectra are characterized by prominent emission of both [Fe~II]
and H$_2$ transitions. The intensity of the [Fe~II]  lines decreases when
moving away from the driving source. In addition to the [Fe~II]  and H$_2$  lines,
emission from other species such as [C~I], [S~II], [N~I], as well as recombination
lines from the Paschen series are detected. These lines are used by \citet{Nisi02}
as a reference to infer the gas-phase iron abundance in the observed HH objects.

In a detailed near-IR spectroscopic study \citet{OConn04} found that, while the CO
emission of the bow shocks in the L\,1634 protostellar outflow originates from cloud
gas directly set in motion, the H$_2$ emission is generated from shocks sweeping
through an outflow. Considering optical data, O'Connell et al. reached a similar
conclusion of a global outflow model involving episodic, slow-precessing, twin jets.

The IR source LDN~1634~7, named IRS~7 by \citet{HoLa95} and located 50'' East
of IRAS~05173$-$0555 \citep[see Figure~\ref{L1634outflow} and][]{HoLa95, Davis97}, may
be associated with a second independent outflow, which extends towards northwest and
southeast (see Figure~\ref{L1634outflow}). The jet associated with LDN~1634~7 seems to
have only two knotty bow shocks \citep[knots 9 and 4 by][]{HoLa95}. \citet{Sea08} reported
Spitzer data of LDN~1634~7 and IRAS~05173$-$0555.

% ----------------------------------------------------------------------------------
% ----------------------------------------------------------------------------------

\section{L1616 and L\,1615}
\label{L1616}

\begin{figure}[!ht]
% \plotfiddle{ngc1788_binned_contrasted.eps}{10.5cm}{0}{40}{40}{-160}{0}
\begin{center}
\includegraphics[width=\textwidth]{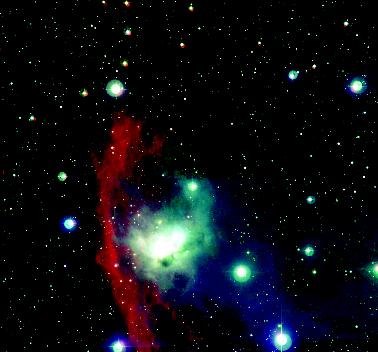}
\caption{Three-color mosaic of L\,1616 and L\,1615 produced by the authors by
combining images taken in the $R$, $I$, and H$\alpha$ filters. The images
were acquired using the Wide Field Imager on the ESO-MPI 2.2m telescope
at La Silla.
The assigned colors are blue, yellow and red for the $R$, $I$, and H$\alpha$
bands respectively. The image covers a field of about 30'$\times$30'.
North is up and East to the left. H$\alpha$ nebular emission
produced by the impact of the hot stars in the OB association to the
north-east appears as an almost vertical "red" lane.
The NGC~1788 reflection nebula is clearly seen to the right of
the H$\alpha$ lane.}
\label{ngc1788}
\end{center}
\end{figure}

\begin{figure}[!ht]
% \plotfiddle{L1616.ps}{8.5cm}{0}{70}{70}{-230}{-160}
\begin{center}
\includegraphics[width=\textwidth]{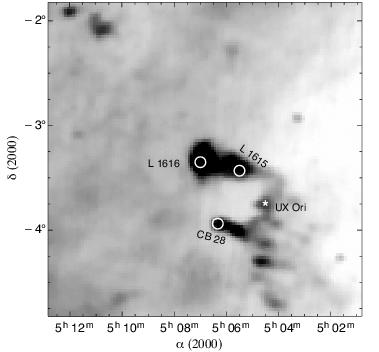}
\caption{Map of the 100~$\mu$m IRAS dust emission of the region around the
L\,1616 cometary cloud. The positions of L\,1615, L\,1616, and CB~28
as reported in Table~\ref{tab:coordinates} are indicated with
circles. The asterisk represents the star UX~Ori.}
\label{clouds_L1616}
\end{center}
\end{figure}

The clouds L\,1616 and L\,1615 \citep{Lyn62} form a cometary cloud
located about 6 degrees west of the Orion giant molecular clouds
(see Figure~\ref{clouds}). The pair of clouds subtends about 40'
(5.2~pc at a distance of 450~pc) roughly in the east-west direction.
The head (L\,1616), pointing toward east, in the general direction of
the Orion OB associations (see Figures~\ref{clouds} and \ref{clouds_L1616}),
harbors the NGC~1788 reflection nebula, which is illuminated mainly
by the B9V-type star HD~293815. A three-color image of L\,1616 and L\,1615
%, produced by combining data obtained with the Wide-Field Imager
% at the ESO 2.2m telescope,
is provided in Figure~\ref{ngc1788}.
L\,1616 and L\,1615 are apparently shaped by the winds and radiation
coming from the massive, hot stars of the OB association.

%%%%%%%%%%%%%%%%%% Fig 5 %%%%%%%%%%%%%%%%%%%%
%%%%%%%%%%%%%%%%%%%%%%%%%%%%%%%%%%%%%%%%%%%%%%%

\citet{Ram95} performed a CO survey in the head of the L\,1616 cometary
cloud and \citet{Stank02} reported on mid-infrared (MIR) and millimeter
observations of the L\,1616 region, which revealed five MIR sources,
associated with very young stellar objects, and four millimeter sources,
the brightest of which may be a Class-0 protostar that drives a powerful
jet.  The data of \citet{Stank02} revealed traces of ongoing star formation
in the cloud and in the NGC~1788 reflection nebula in the head of the
cometary cloud. The location of the protostar discovered by \citet{Stank02}
with respect to other young stars in L\,1616 and %pointing
towards
the OB association suggests an age sequence due to a wave of star formation
propagating through the cloud and triggered by the impact of the nearby OB
association.

%%%%%%%%%%%%%%%%%% Fig  6 %%%%%%%%%%%%%%%%%%%%
%%%%%%%%%%%%%%%%%%%%%%%%%%%%%%%%%%%%%%%%%%%%%%%

A conspicuous X-ray clump was detected in L\,1616 and L\,1615
\citep{Sterz95} which suggested the existence of a small star forming
region. Prior to the RASS, the only known young star in the region
was the T~Tauri star LkH$\alpha$~333. This triggered subsequent X-ray
observations using the High-Resolution Imager onboard ROSAT;
\citet{Alc04}  performed a multi-wavelength study of the L\,1616 region,
from X-ray to near-IR wavelengths. They found more than 20 new low-mass
PMS stars distributed mainly to the east of L\,1616 in about a 1-square-degree
field. They also found that the X-ray properties of the PMS stars in L1616
are quite similar to those of PMS stars detected in the Orion Nebula Cluster.
They derived the stellar parameters for 32 stars in the region and, based on
the level of X-ray emission, lithium content and kinematics, they confirmed
the PMS nature of these stars, as well as their association with L\,1616.
By comparing the position of these stars in the HR diagram with PMS evolutionary
tracks, they inferred an age of about 1-2~Myr with a dispersion of about 1~Myr.
The small age dispersion can be explained in terms of efficient and very
rapid star formation.

Based on CO observations, \citet{Ram95} found an excess velocity of about
1.5~km~s$^{-1}$ above the virial equilibrium velocity of the cloud,
which implies that the virial mass of the cloud is about five times larger
than the observed value. \citet{Ram95} hence concluded that the energy input
due to the stars in the cluster is fragmenting the cloud and, given its size
of about 2~pc, the excess motions of the gas above the virial equilibrium
suggest that the fragmentation process may have lasted for the past 1-2~Myr.
The latter figure is consistent with the age derived by \citet{Alc04} for
the L\,1616 stars.

The most recent cesus of the PMS population of L\,1616/L\,1615 is presented
in \citet{Gan08}. They characterized the young population of the region and
concluded that L\,1616/L\,1615 can be considered as a small cluster according
to the criterion by \citet{Lad03}. \citet{Gan08} also derived the Initial Mass
Function of the region, concluding that it is consistent with that of the field.

Another important issue is the high star formation efficiency  in L\,1616.
From his CO observations, \citet{Ram95} determined a total mass of 180~$M_{\odot}$
for the cloud. The total mass in PMS stars in L\,1616 is at least 30~$M_{\odot}$.
Therefore, the star formation efficiency (mass in stars to total mass fraction) is about
14\% \citep{Ram95, Alc04}, which is significantly high when compared with the
average value of a few percent ($<$3\%), derived for other nearby star forming
regions. The recent investigation by \citet{Gan08} indicates a lower value
(about 8\%) but it is still high in comparison with the canonical values.
% of a few percent.

\begin{figure}[!ht]
% \plotfiddle{L1616_RaDec_flag_Ha.ps}{11cm}{0}{50}{50}{-190}{-10}
\begin{center}
\includegraphics[width=0.9\textwidth]{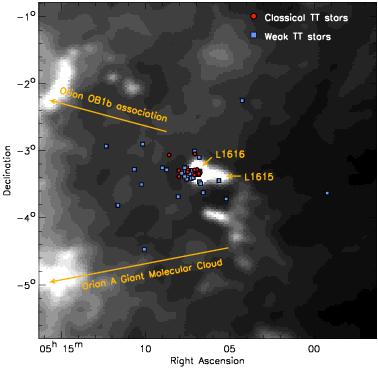}
\caption{Spatial distribution of the PMS stars in L\,1615/L\,1616 overplotted on the
100~$\mu$m IRAS map. The covered area is $5^\circ\times5^\circ$, North is up and
East to the left. The symbols represent the Classical and Weak-line T Tauri stars
according to \citet{Gan08} from which the figure has been adapted here. }
\label{L1616_spadistr}
\end{center}
\end{figure}

Given the spatial distribution of the PMS stars relative to the head of
the cloud (see Figure~\ref{L1616_spadistr}), as well as its cometary shape
and high star formation efficiency, the star formation in L\,1615/L\,1616
may have occurred due to a single triggering event \citep{Alc04, Gan08}.
There is indeed evidence for sequential star formation in the region,
with the Orion OB stars probably being the triggering sources
\citep{Stank02, Gan08}. Possible scenarios for triggered star formation
in L\,1616 have been discussed by \citet{Alc04} and \citet{Gan08}.
Thus, unlike the fossil star forming regions in Orion, L\,1616/L\,1615
appear to be a region of on-going star formation relatively far from
the Orion A and B clouds, in which star formation was most likely
induced by the impact of the massive stars in the Orion OB association.

%%%%%%%%%%%%%%%%%% Fig 7 %%%%%%%%%%%%%%%%%%%%
%%%%%%%%%%%%%%%%%%%%%%%%%%%%%%%%%%%%%%%%%%%%%%%

\subsection{L1616 and L\,1615: PMS Stars}

The most recent and comprehensive list of PMS stars in L\,1616/L\,1615 is
provided by \citet{Gan08}.
% These stars are reported also in Table~\ref{tab:tts}.
The number of confirmed PMS stars in these clouds, including the B9V-type
star HD~293815 and the Herbig Ae/Be star UX~Ori, sums up to 58. The coordinates
and some information on these stars are provided in Table~\ref{tab:tts}, while
more details on their properties can be found in \citet{Gan08}.

The star UX~Ori, which coincides with the IR source IRAS~05020$-$0351,
is located about 45' west of the L\,1615, L\,1616 and CB~28 clouds
(see Figure~\ref{clouds_L1616}). UX~Ori is also the prototype of a class of objects
with similar characteristics and that show similar behavior;
it is an intermediate-mass (M = 2.5M$_{\odot}$)
PMS star with spectral type A3 and age of about 2~Myr \citep{Natta99}.
Its age is consistent with that of the lower mass stars in these clouds
\citep[e.g.][]{Alc04, Gan08}. The star has a strong IR excess which has been
ascribed to a circumstellar disk \citep{Hillen92}. A description of this object
can be found in \citet{Natta99}. UX~Ori shows complex spectroscopic, photometric
and polarimetric variability. This behavior is thought to arise from the fact
that UX~Ori and its circumstellar disk are viewed at high inclination
\citep{Natta99, NatWhit00}. It has also been argued that the observed variability
of UX~Ori may be due to violent comet-like activity
\citep[see][ and references therein]{Grinin01, Beust01, Lagrange00, NatGrinMan00}.
% UX~Ori is the prototype of a group of PMS stars of similar mass, that show
% very similar phenomena.
The near-IR emission of UX~Ori may be due to scattered light if the inner disk
emission is partially obscured by an outer flared disk \citep{Monnier05}.
Extended CO emission, indicating presence of gas distant from the star, has been
detected by \citet{Dent05}.
Such CO emission is possibly associated with the extended far-IR emission
observed by \citet{Natta99}. \citet{Testi01} report millimeter and centimeter
observations of UX~Ori and propose two disk models that can explain
the remarkably flat spectral index observed in the millimeter range.
UX~Ori has been detected by Hipparcos as a possible binary star
\citep{Bertout99}.

\subsection{L1616 and L\,1615: Infrared, Centimeter and Millimeter Sources}

%%%%%%%%%%%%%%%%%%%%%%%%%%%%%%%%%%%%%%%%%%%%%%%%%%%%%%%%%%%%%%%%%%%%%
\begin{table}[p]
\begin{center}
\caption{\label{tab:hhobjects_L1616} Herbig-Haro Objects in L\,1616}

%{\small
\begin{tabular}{lcccl}
\tableline
\noalign{\smallskip}
Cloud & RA (2000) & DEC(2000)          &  cloud       & driving \\
      & h~~~m~~~s & $^{\circ}$~~~'~~~'' &             &  source  \\
\noalign{\smallskip}
\tableline
\noalign{\smallskip}

HH~951-A      & 05:06:39.10 & $-$03:20:46.0 &  L\,1616 & L\,1616~MMS1~A  \\
HH~951-B      & 05:06:49.70 & $-$03:22:23.0 &  L\,1616 & L\,1616~MMS1~A  \\

\noalign{\smallskip}
\tableline
\end{tabular}

\end{center}

\end{table}
%%%%%%%%%%%%%%%%%%%%%%%%%%%%%%%%%%%%%%%%%%%%%%%%%%%%%%%%%%%%%%%%%%%%%

%%%%%%%%%%%%%%%%%% Fig  8 %%%%%%%%%%%%%%%%%%%%
\begin{figure}[p]
% \plotfiddle{L1616_outflow.eps}{6cm}{0}{42}{42}{-185}{0}
\begin{center}
\includegraphics[width=0.95\textwidth]{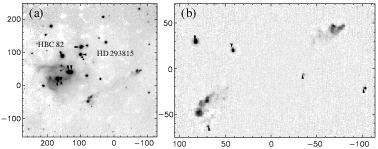}
\caption{Infrared images of the outflow un L1616.
Left panel: 2.12~$\mu$m image of the core of L\,1616; diffuse
H$_2$ emission is clearly seen just below the T Tauri star HBC~82
(=LkH$\alpha$~333). Besides HBC~82 and HD~293815, the other arrows
indicate the five mid-IR sources detected by \citet{Stank02}.
Right panel: enlargement of the region of the outflow. The arrows
indicate stellar sources. The offsets are in arc-seconds from the
position of the millimeter source L\,1616~MMS1~A. The components
of both HH~951-A and HH~951-B can be appreciated.
Adapted from \citet{Stank02}. }
\label{L1616outflow}
\end{center}
\end{figure}
%%%%%%%%%%%%%%%%%%%%%%%%%%%%%%%%%%%%%%%%%%%%%%%

%%%%%%%%%%%%%%%%%% Fig 9 %%%%%%%%%%%%%%%%%%%%
\begin{figure}[p]
% \plotfiddle{L1616_jet_Ha.ps}{5cm}{0}{43}{43}{-45}{-124}
% \plotfiddle{ngc1788_detail3.ps}{0cm}{0}{25.5}{25.5}{-180}{-30}
\begin{center}
\includegraphics[height=0.3\textwidth]{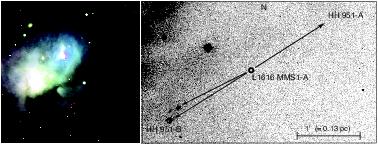}
\caption{Optical images of the outflow in L1616.
Left panel: Detail of the central part of the L\,1616
image shown in Figure~\ref{ngc1788}. It covers approximately
the same field of view as in Figure~\ref{L1616outflow}~(a),
i.e. about 6.7'$\times$6.7'. North is up and East to the left.
The H$\alpha$  emission from the components of HH~951-B can be
appreciated as the small red cloud fragments just below the
center of the field. Right panel: a $3.4'\times2'$ detail of
the WFI H$\alpha$ image of the outflow in L\,1616. The HH~951-B
component can be clearly appreciated in the lower left.
The position of the driving source is marked with the circle.
HH~951-A is bitten by the high optical extinction to the west
of NGC~1788.}
\label{L1616_jet_Ha}
\end{center}
\end{figure}
%%%%%%%%%%%%%%%%%%%%%%%%%%%%%%%%%%%%%%%%%%%%%%%

% ----------------------------------------------------------------------------------
% ----------------------------------------------------------------------------------

One of the most interesting IR sources in this region is IRAS~05020$-$0351,
which coincides with the Herbig Ae/Be star UX~Ori. Another one is IRAS~05044$-$0325,
which can be identified with the H$\alpha$-emission star Kiso~A-0974~15.
The other source is IRAS~05076$-$0257, that can be identified with the star
V1011~Ori, which turned out to be a T~Tauri star \citep{Alc04}.
\citet{Stank02} reported mid-infrared and 1.2~mm observations in L\,1616, which
revealed five MIR sources, that are associated with YSOs. Their coordinates
are reported in Table~\ref{tab:tts}. The 1.2~mm observations revealed a group
of four dust continuum sources, whose coordinates are also reported in
Table~\ref{tab:tts}. The brightest millimeter source, which drives a near-IR
H$_2$ powerful jet (see Section~3.3), may be a Class-0 protostar
\citep{Stank02}. In addition, there are three IRAS sources close to CB~28,
namely, IRAS~05038$-$0400, IRAS~05036$-$0359 and IRAS~05037$-$0402.
The H$\alpha$-emission star Kiso~A-0974~13 is also located very close
to this cloud.

\subsection{L1616 and L\,1615: Outflows and Herbig-Haro Objects}
\label{outflow_L1616}

The near-IR imaging by \citet{Stank02} in the core of L\,1616 led to the
identification of three features of H$_2$ line emission.
One of these features can be appreciated as diffuse emission that may be
related to the reflection nebulosity surrounding
the brightest members of the cloud (see Figure~\ref{L1616outflow}, left panel).
As Stanke et al. mention, it is most likely that this feature is due to
fluorescent emission from UV-excited H$_2$ in the vicinity of the most
massive members of L\,1616, in particular the B9~V star HD~293815.

The other two features can be clearly identified with two bow-shock
structures north-west and south-east of the millimeter sources
reveled in this region (cf. Figure~\ref{L1616outflow}).
They are apparently shocks due to a protostellar outflow driven
by one of the millimeter sources. Stanke et al. consider the most
massive millimeter source, L\,1616~MMS1-A, as the most likely driving
source of the outflow. The projected length of the outflow is about 190'',
which corresponds to  $\sim$0.41 pc at a distance of 450~pc.
Stanke et al. did not find any further emission beyond the two bow shocks.
The approximate coordinates of the brightest knots in this outflow are
reported in Table~\ref{tab:hhobjects_L1616}. This object has been assigned
the number HH~951, following the criteria for the HH numbers designation
\citep{Reip00}. HH~951-B is detected in H$\alpha$ images, but HH~951-A is
barely seeing in such images (see Figure~\ref{L1616_jet_Ha}).
In addition, [S~II] emission from HH~951 has been detected on an objective
prism plate (Reipurth, private communication).

\section{IC~2118}
\label{IC2118}

\begin{figure}[!ht]
% \plotfiddle{IC2118_Three_Color.ps}{10cm}{0}{65}{65}{-190}{-130}
\begin{center}
\includegraphics[width=0.9\textwidth]{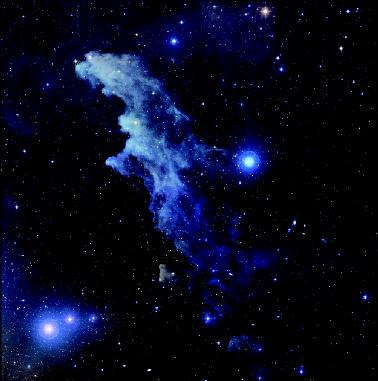}
\caption{Optical color image of the IC~2118 region, adapted from
an image by Noel Carboni. The image covers a field of about
$4^\circ \times 4^\circ$; North is up and East to the left.}
\label{IC2118_color}
\end{center}
\end{figure}

\begin{figure}[!ht]
% \plotfiddle{ic2118.ps}{10.5cm}{0}{70}{70}{-240}{-125}
\begin{center}
\includegraphics[width=0.9\textwidth]{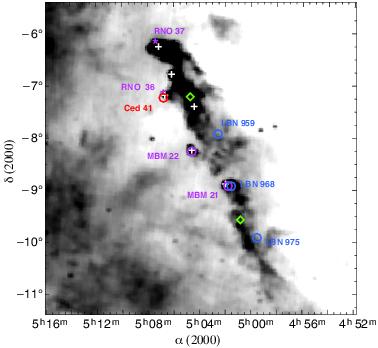}
\caption{Map of the 100~$\mu$m IRAS dust emission in the IC~2118 region.
The positions of the clouds according to Table~\ref{tab:ic2118clouds}
are indicated with circles.
The positions of the RNOs reported by \citet{Cohen80} are marked with
asterisks, while the clouds detected in the CO survey by \citet{Kun01}
are represented with white {\it plus} symbols.
The southern and northern diamonds mark the positions of the two reflection
nebulae, DG~49  and DG~52 respectively, listed by \citet{DG66}.
}
\label{clouds_IC2118}
\end{center}
\end{figure}

The IC~2118 region, also known as the {\it witch-head nebula}, is a long,
filamentary reflection nebula, located at an angular distance of about
8 degrees west of the Orion~A molecular cloud (cf. Figure~\ref{clouds}
and \ref{clouds_IC2118}). An optical color image of the cloud is shown
in Figure~\ref{IC2118_color}.
The cloud is also known as Ced~41 \citep{Ceder46} and Hubble~9 \citep[][catalog]{Hu22}.
IC~2118 lies far from the Orion OB1 association, clearly outside the region
occupied by luminous stars, and also well separated from the Orion main
molecular clouds. The supergiant star {\it Rigel} ($\beta$~Ori), seen
projected at about two degrees east of IC~2118, is thought to be the
illuminating source of the nebula.
A good description of the IC~2118 region and its associated clouds
can be found in \citet{Kun01}.

%%%%%%%%%%%%%%%%%% Fig  10 %%%%%%%%%%%%%%%%%%%%
%%%%%%%%%%%%%%%%%%%%%%%%%%%%%%%%%%%%%%%%%%%%%%%

A number of clouds and reflection nebulae are found in this region, whose
positions are marked in the 100~$\mu$m IRAS map of Figure~\ref{clouds_IC2118}
while their coordinates and designations are listed in Table~\ref{tab:ic2118clouds}.
\citet{DG66} reported two reflection nebulae in this region: DG~49 and DG~52,
\citep{DG63}, marked by the diamonds in Figure~\ref{clouds_IC2118}, and which
indicate the two bigest portions of the cloud. The \citet{Lyn65} catalogue
contains three associated bright nebulae, namely,  LBN~968, LBN~959 and LBN~975,
that can be considered as sub-structures of the same cloud.
\citet{Cohen80} found two red nebulous objects (RNO) in the region,
namely RNO~36 and  RNO~37. In his Table~1 he describes RNO~36 as two
extremely faint, very red stars located on an emission-rim with very faint
general background nebulosity present, and RNO~37 as a group of about 18
very faint, very red stars, where the dominant two are nebulous stars.
Cohen performed spectroscopic observations for seven of the stars in the
RNO~37 group, most of them presenting spectral types later than K5 and
emission lines.

%%%%%%%%%%%%%%%%%% Fig  11 %%%%%%%%%%%%%%%%%%%%
%%%%%%%%%%%%%%%%%%%%%%%%%%%%%%%%%%%%%%%%%%%%%%%

\begin{table}[!ht]
%% S.Rodney 071017 : reduce table column separation to squeeze whitespace
\setlength{\tabcolsep}{0.5\tabcolsep}
\caption{\label{tab:ic2118clouds} Coordinates of the Dark Clouds and Bright Nebulae in the IC~2118 Region.}
% \label{ic2118clouds}
% \smallskip
\begin{center}
{\small
\begin{tabular}{l@{\hskip8pt}c@{\hskip5pt}c@{\hskip8pt}c@{\hskip6pt}c@{\hskip8pt}l@{\hskip8pt}l}
\tableline
\noalign{\smallskip}
Cloud & RA (2000) & DEC(2000)           & $l$       & $b$ & other IDs &  refs. \\
      & h~~~m~~~s & $^{\circ}$~~~'~~~'' & ($^\circ$) & ($^\circ$) &         &         \\
\noalign{\smallskip}
\tableline
\noalign{\smallskip}
G~206.4$-$26.0 & 05:07:11 & $-$06:15:06 & 206.400  & $-$26.000 &                    & 1       \\
G~206.8$-$26.5 & 05:06:10 & $-$06:47:01 & 206.800  & $-$26.467 &                    & 1       \\
G~207.3$-$26.5 & 05:06:41 & $-$07:11:11 & 207.267  & $-$26.533 & Ced~41, Hubble~9   & 1, 2, 3 \\
G~207.2$-$27.1 & 05:04:25 & $-$07:24:11 & 207.200  & $-$27.133 &                    & 1       \\
G~208.1$-$27.5 & 05:04:36 & $-$08:14:30 & 208.067  & $-$27.467 & MBM~22             & 1, 4    \\
G~208.4$-$28.3 & 05:01:59 & $-$08:53:18 & 208.400  & $-$28.333 & MBM~21             & 1, 4    \\
LBN~959        & 05:02:34 & $-$07:55:53 & 207.500  & $-$27.780 &                    & 5       \\
LBN~968        & 05:01:33 & $-$08:55:55 & 208.390  & $-$28.450 &                    & 5       \\
LBN~975        & 04:59:32 & $-$09:55:21 & 209.160  & $-$29.330 &                    & 5       \\

\noalign{\smallskip}
\tableline
\noalign{\smallskip}

\multicolumn{7}{l}{\parbox{0.95\textwidth}{\footnotesize
 References:
1:~\citet{Kun01};
2:~\citet{Ceder46};
3:~\citet{Hu22};
4:~\citet{MagBliMun85};
5:~ \citet{Lyn65}
 }} \\

\end{tabular}
}

\end{center}
\end{table}

In a survey for high-latitude molecular clouds \citet{MagBliMun85}
detected two molecular clouds in the region of IC~2118: MBM~21 and MBM~22.
\citet{Ogu98} classified the clouds associated with IC~2118 as remnant
molecular clouds. The clouds MBM~21 and MBM~22 were included in the IR
survey by \citet{reach98}.
Using the NANTEN telescope \citet{Yon99} investigated the molecular clouds
associated with the IR sources IRAS~04591$-$0856 (G208.3$-$28.4) and
IRAS~05050$-$0614 (G206.4$-$25.9) in the $^{12}$CO, $^{13}$CO, and C$^{18}$O
transitions.
The molecular clouds G208.3$-$28.4 and G206.4$-$25.9 detected by
\linebreak[4] \citet{Yon99}
%Yonekura et al.
coincide spatially with MBM~21, investigated by \citet{MagBliMun85},
and RNO~37 \citep{Cohen80}, respectively.
The IRAS sources have infrared flux density distributions characteristic
of YSOs, and their association with molecular clouds is a strong indication
of star formation in those clouds.
The $^{12}$CO maps show that MBM~21 has a cometary shape with the head
pointing towards the Orion OB1 association, while the $^{13}$CO data
indicate that such cometary structure is formed by photoevaporation \citep{Yon99}.

\citet{Kun01} performed a $^{12}$CO survey in about 6 square-degrees
in the region of IC~2118 and identified six molecular clouds, among
which MBM~21 and MBM~22, as well as the other clouds discussed above,
and derived the physical properties of the clouds.
In Table~\ref{tab:ic2118clouds} the coordinates of the clouds in the
region of the reflection nebula IC~2118 are provided, together with other
cloud designations, while Figure~\ref{clouds_IC2118} shows the clouds on
the 100$\mu$m IRAS dust emission map.

\subsection{IC~2118: PMS Stars}

Based on an objective-prism survey, \citet{Kun01} identified 46 candidate
emission-line stars which were more recently investigated spectroscopically
by \citet{Kun04}. They identified five classical T Tauri stars in the region
of IC~2118 which are listed in Table~\ref{tab:tts}. Three of them were previously
known IRAS sources. Using the near-IR magnitudes from the 2MASS catalogue and
adopting a distance of 210~pc (see Section~\ref{distances}), \citet{Kun04} estimated,
by comparison with \citet{PaSta99} PMS evolutionary tracks, an average age of
2.5~Myrs and masses in the interval 0.4-1.0~M$_{\odot}$ for the members of
the IC~2118 association. They concluded that the five classical T~Tauri stars
projected on the clouds are physically related to them, and that star formation
in the region was most likely triggered by shock waves possibly originating
from the Orion~OB1 association. In addition, \citet{Kun04} report other three
stars that they indicate as PMS star candidates. Two of these (2MASS\,J\,05060574$-$0646151
and 050944864$-$0906065) are most likely weak-line T~Tauri stars, while the
other one (2MASS\,J\,05112460$-$0818320), showing very strong variations in its
spectrum, may be a classical T~Tauri star.
One of the two weak-line T~Tauri star candidates (2MASS\,J\,05060574$-$0646151)
is projected on the molecular cloud G~206.4-26.0, while the other and the
classical T~Tauri star candidate are located far from the IC~2118 clouds.
These three stars are also reported in Table~\ref{tab:tts} as T~Tauri star
candidates in IC~2118.

Other four T~Tauri stars in the field of IC~2118 were identified as optical
counterparts of ROSAT All-Sky Survey sources and their properties are reported
in \citet{Alc98, Alc00}.
Although these objects are projected relatively far from the IC~2118 clouds
we also include them in Table~\ref{tab:tts} as possible members of IC~2118.

\subsection{IC~2118: Infrared, Centimeter and Millimeter Sources}

Early studies revealed the IR sources IRAS~05050$-$0614 and IRAS~04591$-$0856,
both identified with young stars.  IRAS~05050$-$0614 is located in the head of
the northernmost part of the nebula. It is identified with the H$\alpha$-emission
star Kiso~A-0974~19 and was later confirmed to be a classical T~Tauri star by
\citet{Kun04} (see Table~\ref{tab:tts}). More controversial was the identification
of IRAS~04591$-$0856. Being located in the MBM~21 cloud, it coincides with a small
nebulous object catalogued as a Herbig--Haro object (G~13 by \citet{Gyu82};
and HHL~17 by \citet{Gyu87}). However, its Herbig--Haro nature was
not confirmed later.
\citet{Persi88} found a spectral energy distribution that is intermediate
between that of a low-mass protostar and a deeply embedded T~Tauri star.
Analogous conclusion was reached by \citet{Tap97}, who found that
IRAS~04591$-$0856 is a heavily reddened (A$_V$ $<$ 15 mag) T~Tauri star,
but did not detect any nebulosity around the very red point-like object
in their near-IR images.
\citet{Kun01} found that the spectral energy distribution of IRAS~04591$-$0856
in the 1.25--100~$\mu$m wavelength range is characteristic of Class-I sources,
while the optical visibility suggests a more evolved nature. Finding strong
variability, \citet{Kun01} concluded that IRAS~04591$-$0856 is most likely
a low-mass star near the birthline.
In the 2MASS images this object appears as point-like but with a surrounding
nebulosity. Thus, we list this source in Table~\ref{tab:tts} as a low-mass
PMS object.

Additional objects in IC~2118 come from \citet{Kun01} who selected
11 faint IR sources from the IRAS Point Source Catalogue and Faint Source
Catalogue in the region.
Three of those, namely IRAS~04587$-$0854, IRAS~F~05044$-$0621 and IRAS~F~05047$-$0618,
have a 2MASS counterpart; the latter two were found to be classical T~Tauri stars
in subsequent spectroscopic follow-up observations by \citet{Kun04} (see Table~\ref{tab:tts}).
The remaining IR sources were found to be parts of small but extended structures
at 12 and 25~$\mu$m, though most of them appeared to be point-like at
60 and 100~$\mu$m, suggesting that they are small-scale density or
temperature enhancements in the cloud with sizes of 5'--10', corresponding
to 0.3--0.6 pc at a distance of 210~pc \citep{Kun01}.

\section{L~1642}
\label{L1642}
L~1642 is a well known high Galactic latitude translucent/dark cloud,
also listed as MBM~20 in \citet{MagBliMun85}. Based on previous multiwavelength
data \citep{Tay82, Lau87, Lilje88, Lilje91}, L\,1642 appears as a cool and
quiescent cloud, associated with a much larger HI cloud (over 4 deg) with
cometary structure. The tail, extending more than 5 degrees in the north-east
direction, is perpendicular to the Galactic plane and points towards the
plane. On the sky, L\,1642 is projected in the direction of the edge of the
Orion-Eridanus Bubble, with which it might be interacting \citep{Leht04}.

In general, high latitude molecular clouds are considered young and transient
structures \citep{Magn93,Heith96} and only in rare cases these clouds show evidence
of star formation.
L~1642 is one of the two high-latitude ($\|b\|>30^\circ$) clouds known to have
star formation \citep[the other one is MBM~12, see][and references therein]{Luhm01}.

The morphology and physical properties of L\,1642 have been recently investigated
by \citet{Russeil03} in the CO, and by \citet{reach98}, \citet{Vert00} and
\citet{Leht04, Leht07} in the mid- and far-IR. A detailed study of the optical
and IR properties of the dust grains in the region is reported in \citet{Leht07}.
The maximum optical extinction in the central part of the cloud has been estimated
to be A$_v=8$ mag, based on near-IR color excesses derived from 2MASS data, while
an extended lower extinction halo is seen around the central core from extinction,
optical scattered-light and far-IR 100~$\mu$m IRAS maps \citep{Leht04, Leht07}.
Large-scale mapping at 200~$\mu$m with ISO showed that the cloud consists of
three separate denser regions connected by diffuse material \citep{Leht04, Leht07}.
Two filaments of dust extend respectively to the North-East and East from the main cloud.
Only the region with the highest dust column density corresponds to a temperature
minimum of 13.8~K. Close to this densest core there are two nebulous low-luminosity
PMS binary stars, L\,1642-1 (EW~Eri) and L\,1642-2 (HBC~410) both detected by IRAS
\citep{Sand87}. \citet{Leht04} estimate that the ratio between virialized and
observed mass is about 1.2 for the densest part (region  B), while for other regions
it is about 10 and conclude that, within uncertainties, only region B can be gravitationally
bound, while the other parts of L\,1642 are transient \citep[see Fig.~2 in][]{Leht04}.
A decrease in dust temperature towards the center of the dense region B, by an
amount that cannot be explained only in terms of attenuation of the radiation
field, and an increase in apparent emissivity in the colder regions are observed
in the cloud; these two phenomena can be explained in terms of an increase of
the dust absorption cross-section at far IR wavelengths \citep{Leht07}.

\subsection{L~1642: PMS Stars}
There are two PMS binaries in this region: L\,1642-1 and L\,1642-2.
L~1642-1A is optically identified as a K~7IV T~Tauri star, obscured by
about 2~mag of visual extinction \citep{Sand87}.
The secondary, at 2.7~arsec separation, appears quite bright in the far-red,
and might become dominant in the IR.
L~1642-2A, a faint M0 H$\alpha$-emission star associated with a small
compact reflection nebula, is the powering source of the Herbig-Haro object
HH~123 \citep{ReipHeat90}. Its secondary component, at about 5.4~arcsec
separation, is also a H$\alpha$-emission star, has a redder color than the
primary, and might be responsible for the far-IR emission.
% This is also a H$\alpha$-emission star and is redder than the primary.
The two components correspond altogether to the IR source IRAS~04325$-$1419.
A small, weak molecular outflow is centered on this object \citep{Lilje89}.
This source was also included in the 1300~$\mu$m survey by \citet{Reipetal93}
who derived a bolometric luminosity of 0.4~L$_\odot$.
Both binaries are projected on regions of high extinction, and have very low
luminosity (less than about 0.5~M$_{\odot}$, if a distance of 100~pc is adopted).
Both secondaries appear very active:
L~1642-1B shows Br-$\gamma$ emission, while L\,1642-2B exhibits H$_2$ emission
in the IR \citep{Sand87}.

\subsection{L~1642: Infrared, Centimeter and Millimeter Sources}
Apart from L\,1642-1, and L\,1642-2, four additional IRAS Point Source Catalogue
objects also fall within the boundaries of the 200~$\mu$m map of \citet{Leht04},
e.g.: IRAS~04336$-$1412 (L~1642-3), IRAS~04347$-$1415 (L~1642-4), IRAS~04342$-$1444
(BD-14 929) and IRAS~04349$-$1436, which are detected only at 60~$\mu$m,
100~$\mu$m, 12~$\mu$m and 100~$\mu$m, respectively. None of the IRAS sources,
nor any other point source candidates, is detected at 200~$\mu$m.

\subsection{L~1642: Outflows and Herbig-Haro Objects}
This cloud hosts the Herbig-Haro object HH~123 ($\alpha_{2000}=04^h34^m49.60^s$;
$\delta_{2000}=-14^\circ13'08{\farcs}0$), discovered by \citet{ReipHeat90}, who
described it as a slightly elongated amorphous object. From their low-dispersion
spectra they found that the object is of intermediate excitation with a reddening
$E(B-V) \approx0.4$. From their high-resolution spectra a number of red lines are
detected which reveal three emission components with velocities of about -70, 0,
and +110~km~s$^{-1}$, respectively. They interpreted these components in terms
of two bow shocks moving away, in opposite directions, from the source L\,1642-2,
which is driving the outflow.

\section{Other Small Orion Outlying Clouds}
\label{other_clouds}

Other Orion outlying clouds with much less information in the literature that
we discuss briefly here are the two small clouds CB~28 and CB~29 coming from
the catalogue of optically selected clouds by \citet{Clem88},  and the
Lynds bright nebulae LBN~991, LBN~917, and LBN~906 \citep{Lyn65}.

\subsection{CB~29}
CB~29 is a small ($\sim$18'; or $\sim$2~pc at the distance of 460~pc)
cloud that lies about 3 degrees to the west of the Orion OB1a association
and about 40' to the north of L\,1634 (see Figures~\ref{clouds} and
\ref{clouds_L1634}). The properties of this cloud have been studied by
\citet{Clem88}. Four IRAS sources are found closeby (e.g. IRAS~05194$-$0343,
IRAS~05194$-$0346, IRAS~05201$-$0341 and IRAS~05190$-$0348), while three
H$\alpha$ emission  stars (e.g. Kiso~A-0975~65, Kiso~A-0975~66 and
Kiso~A-0975~67) appear projected on the cloud.

\subsection{CB~28}
This is the small cloud located some 25' south of the L\,1616 cometary
cloud (see Figure~\ref{clouds_L1616}).
This cloud coincides with the Lynds bright nebula LBN~923, and the object
No.~67 catalogued by \citet{Lee99}, which is considered to be a starless core.
The CB~28 cloud has been studied in the IR by \citet{reach98}
and in CO by \citet{Park04}. The radial velocity of the cloud, as
determined from the $^{12}$CO and $^{13}$CO lines reported by \citet{Park04},
is consistent with that of the Orion giant molecular cloud.

\subsection{LBN~991, LBN~917 and LBN~906}

LBN~991 is located at about 8 degrees in angular distance
south-west of the Orion~A molecular cloud. % There is
Not much investigation has been carried out on this cloud but we % have
decided to include it in this chapter because it coincides with one of the X-ray
clumps detected in the analysis of the ROSAT All-Sky Survey X-ray sources in Orion.
In the 100~$\mu$m IRAS map LBN~991 appears as a diffuse cloud (see Figure~\ref{clouds}).

LBN~917 and LBN~906 are some of the most distant clouds, in angular distance,
from the Orion association that are discussed here (see Figure~\ref{clouds}).
LBN~917 was detected  in the IR survey by
\citet{reach98}, who identified the clouds as DIR~203$-$32. \citet{Bally91}
concluded that these clouds must have been either ejected from a region near
the Orion main molecular clouds or condensed from the expanding HI shell
surrounding the Orion clouds. These clouds are indeed projected at an angular
distance of more than 10 degrees from the Orion giant molecular cloud and are
more probably related to the Orion-Eridanus bubble.
Further CO observations of LBN~917 were performed by \citet{Magn00} and
\citet{Onish01}.  Some scattered T Tauri and H$\alpha$ emission-line stars
are found in the surroundings. Their coordinates are reported in
Table~\ref{tab:tts}.

The IRAS source IRAS~04451$-$0539, close to LBN~917 and LBN~906, was classified
as a T~Tauri star by \citet{GrHe92}, based on H$\alpha$ emission and strong
lithium $\lambda$6708~\AA~ absorption in its optical spectrum.

On the other hand, several RASS sources form an X-ray clump close to
the LBN 991 nebula (see Figure~\ref{clouds}). The X-ray source
RX~J0513.4$-$1244, classified
as a G4 weak-line T~Tauri star by \citet{Alc00}, and the H$\alpha$-emission star
Kiso~A-1047-1 are projected in the proximity of the nebula, though the IR colors
of the latter do not satisfy the criteria adopted by \citet{Lee05}. Therefore,
there might be many more X-ray emitting PMS stars still to be discovered in
this region.

\subsection{Anonymous X-ray Clump}
A concentration of X-ray sources that satify the criteria by \citet{Sterz95}
was detected at $\alpha \approx 05^h21^m$, $\delta \approx -09^\circ$, some
3~degrees to the south of L\,1634 (see Figure~\ref{clouds}).
Several PMS stars are found in the vicinity of this clump:
three are identified as optical counterparts of ROSAT All-Sky Survey
sources \citep{Alc96, Alc00} and one is an IRAS source (IRAS~05222$-$0844)
classified as a T~Tauri star by \citet{GrHe02}. The latter was also detected
in X-rays with ROSAT \citep{Alc96, Alc00}.
In addition, 9 Kiso H$\alpha$-emission stars are found around the X-ray
clump, but we do not include them in Table~\ref{tab:tts} because their
IR (2MASS) colors do not satisfy the criteria for PMS stars indicated
by \citet{Lee05}. Follow-up spectroscopy is needed to asses their PMS
nature.

\section{Small Globules around $\sigma$~Ori}

\begin{figure}[!ht]
% \plotfiddle{glob_dss.ps}{11cm}{0}{70}{70}{-210}{-110}
\begin{center}
\includegraphics[width=0.9\textwidth]{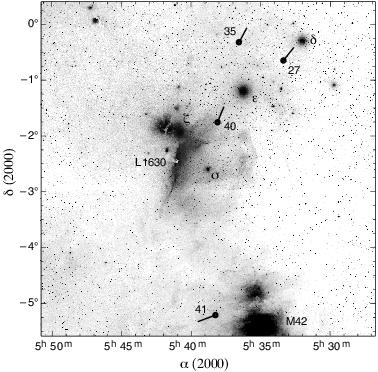}
\caption{Spatial distribution of the small globules
around $\sigma$~Ori discussed in the text overplotted on
the red digitized sky survey image. $\sigma$~Ori is at
the center of the image. The Orion belt stars are identified
by the greek letters while the numbers indicate the globules
according to the designation by \citet{Ogu98}.
Their positions are marked by black dots with pointers
indicating the direction of their tails.
The Orion nebula (M42) and the Horsehead nebula (L1630)
are also indicated.}
\label{sOri_glob_tails}
\end{center}
\end{figure}

\begin{figure}[!ht]
% \plotfiddle{OS98-27_150bpi.ps}{10.5cm}{0}{35}{35}{-200}{70}
% \plotfiddle{OS98-35_150bpi.ps}{0cm}{0}{35}{35}{-8}{94}
% \plotfiddle{OS98-40_150bpi.ps}{0cm}{0}{35}{35}{-200}{-70}
% \plotfiddle{OS98-41_150bpi.ps}{0cm}{0}{35}{35}{-8}{-46}
\begin{center}
\includegraphics[width=\textwidth]{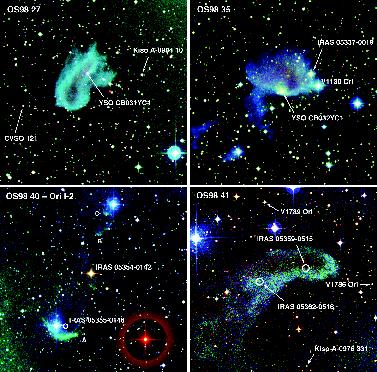}

\caption{Optical color images of the four small globules around
$\sigma$-Ori dicussed in the text. The images were produced by
the authors using the publicly available blue, red and IR digitized
sky survey plates. Some of the most relevant objects are indicated and
the position of some of the IRAS sources are marked with circles.
Each image covers a field of about 15'$\times$15', with North up
and East to the left.}
\label{sOri_globules}
\end{center}
\end{figure}

One of the most massive stars in Orion is the O7V star $\sigma$~Ori.
For a review on the cluster around this object we refer the reader to
the chapter by Walter et al. in this book. In this section we
discuss the four small globules No. 27, 35, 40 and 41 by \citet{Ogu98},
located in the vicinity of $\sigma$~Ori, and whose morphology and
characteristics suggest a possible evolutionary sequence of remmant clouds,
starting from bright-rimmed clouds, through cometary globules, to reflection
clouds, as proposed by \citet{Ogu98}.
The coordinates of these globules, their approximate size and designations
are listed in Table~\ref{tab:sig_ori_globs}, while their spatial distribution
relative to $\sigma$~Ori is shown in Figure~\ref{sOri_glob_tails}. Three of
the globules (27, 35 and 40) lie to the north-west of the star, while the
other one (41) is located to the south. From Figure~\ref{sOri_glob_tails}
it is evident that the tails of OS98~27, OS98~35 and OS98~40 point away
from $\sigma$~Ori, which underlines the impact of the strong wind from the
massive star on these globules. There is evidence of recent or ongoing
star formation in these clouds. %A discussion on a cloud by cloud basis
A discussion cloud by cloud is presented below and, as a reference, optical
color images of each globule are shown in Figure~\ref{sOri_globules}.

%%%%%%%%%%%%%%%%%%%%%%%%%%%%%%%%%%%%%%%%%%%%%%  Table %%%%%%%%%%%%%%%%%%%%%%%%%%%%%%%%%%
\begin{table}[!ht]
%% S.Rodney 071017 : reduce table column separation to squeeze whitespace
\setlength{\tabcolsep}{0.5\tabcolsep}

\caption{\label{tab:sig_ori_globs} Small Globules around $\sigma$~Ori
 discussed in this Chapter.}
\begin{center}
{ \small
%\begin{tabular}{lcccll}
\begin{tabular}{l@{\hskip8pt}c@{\hskip6pt}c@{\hskip6pt}c@{\hskip6pt}l@{\hskip8pt}l}
\tableline
\noalign{\smallskip}
OS98$^\dagger$  & $\alpha$(2000) & $\delta$(2000)       &  Approx. Size & other IDs     & refs.  \\
                & h~~~m~~~s      & $^{\circ}$~~~~'~~~~''& (arcmin)      & 	       &        \\
\noalign{\smallskip}
\tableline
\noalign{\smallskip}
27           & 05:33:24      &        $-$00:38:03      &  3$\times$5    &  IC~423, CB~31, LBN~913, DG~58 &  1, 2, 3 \\
35           & 05:36:30      &        $-$00:17:26      &  6$\times$4    &  IC~426, CB~32, LBN~921, DG~61 &  1, 2, 3 \\
40           & 05:38:06      &        $-$01:45:29      &  3$\times$5    &  Ori~I-2, BRC~20      	 &  2, 4, 5 \\
41           & 05:38:24      &        $-$05:13:49      &  3$\times$8    &  BRC~22   	                 &  4       \\
\noalign{\smallskip}
\tableline
\noalign{\smallskip}

\multicolumn{6}{l}{\parbox{0.9\textwidth}{\footnotesize
    $^\dagger$ Cloud number by \citet{Ogu98} }}\\
\multicolumn{6}{l}{\parbox{0.9\textwidth}{\footnotesize
    References:
    1:~\citet{Drey1908}: {\it Index Catalog};
    2:~\citet{Clem88};
    3:~\citet{Lyn62};
    4:~\citet{Sugi91};
    5:~\citet{Cer92};
}}\\

\end{tabular}
}
\end{center}
\end{table}
%%%%%%%%%%%%%%%%%%%%%%%%%%%%%%%%%%%%%%%%%%%%%%%%%%%%%%%%%%%%%%%%%%%%%%%%%%%%%%%%%%%%%%

%%%%%%%%%%%%%%%%%% Fig  12 %%%%%%%%%%%%%%%%%%%%
%%%%%%%%%%%%%%%%%%%%%%%%%%%%%%%%%%%%%%%%%%%%%%%

%%%%%%%%%%%%%%%%%% Fig 13 %%%%%%%%%%%%%%%%%%%%
%%%%%%%%%%%%%%%%%%%%%%%%%%%%%%%%%%%%%%%%%%%%%%%

\subsection{OS98-27}
 This small globule has the typical morphology of a cometary cloud
 with its tail pointing towards north-west.  There is evidence of
 star formation with at least two confirmed classical T~Tauri stars,
 namely CVSO~121 and YSO~CB031YC1. These two objects were found to
 posses strong H$\alpha$ in emission, as well as strong lithium
 $\lambda$6708~\AA~ absorption in their spectrum by \citet{Bri05} for
 the former and by \citet{Yun97} for the latter. While CVSO~121 is
 located $\sim$3~acr-min. to the east of the head of the cometary
 cloud, YSO~CB031YC1 is on the tail of the cloud (see the left upper
 panel of Figure~\ref{sOri_globules}). In addition, the emission line
 star Kiso~A-0904~10, is located to the north-west of the globule.
 These objects are also listed in Table~\ref{tab:tts}.

\subsection{OS98-35}

 Several cloud fragments were identified by \citet{Ogu98} in this region.
 The most prominent is OS98-35A. Although more extended in the east-west
 direction, this globule also shows the typical morphology of the head
 of a cometary cloud. There is also evidence of star formation in this
 cloud: the star YSO~CB032YC1, spectroscopically confirmed to be a T~Tauri
 by \citet{Yun97}, is located at the apex of the cloud rim (see upper right
 panel in Figure~\ref{sOri_globules}). \citet{Yun97} do not provide the
 optical magnitudes for this object, but we estimate V$\approx$14.
 The ROSAT X-ray source 1RXS J053631.1-001744 is about 1~arc-min to the
 north of the T Tauri star.
 The source IRAS~05337$-$0019 is located on the western edge of OS98-35A
 and coincides with the star TYC~4766-2306-1. The latter is at less than
 20 arc-sec from the ROSAT X-ray source 1RXS J053620.0-001708.
 Given the typical error circle of the ROSAT sources, with a radius on
 the order of 30~arc-sec, the star may be associated with the X-ray
 source. However, spectroscopic evidence of the young nature of this
 object is missing. Hence it is listed among candidates in
 Table~\ref{tab:tts}.
 Finally, the unrelated  % $\alpha^2$ CVn-type
 variable star V1130~Ori is also projected on the border of the
 OS98-35 nebulosity.

\subsection{OS98-40}
 This cloud, projected closer to $\sigma$~Ori than OS98-27 and OS98-35
 (cf. Figure~\ref{sOri_glob_tails}), is also very well known as Ori~I-2.
 A detailed study of Ori~I-2, including its kinematics, was condicted by
 \citet{Cer92} using CO observations.
 The three cloud fragments A, B, and C, as identified by \citet{Ogu98},
 are marked on the image of the cloud shown in the lower left panel
 of Figure~\ref{sOri_globules}. The most prominent of the three is fragment A,
 but the "V-shape" morphology of the three cloud fragments puts in evidence
 the interaction of the strong wind from $\sigma$~Ori with the cloud material
 \citep[cf.][]{Cer92}.
 The bright and very red star in the lower right corner of this image is
 the unrelated Mira variable X~Ori.

 A radio source, Ori-I-2~1, detected by \citet{Bot96}, is located just on
 the apex of cloud A. The source IRAS~05355$-$0146 is located in the cavity
 of this cloud and the closeby bright star to the east  is HD~37389,
%%, just to the east of it,
 an unrelated B9-A0V star \citep{Cer92}. The IRAS source has been associated
 with an H$_2$O maser \citep{Wou86, Cod95} and is the driving source of the
 HH~289 outflow (see below).
 Four emission line stars (OriI-2N-1, 2, 3, and 4) in the northern part of
 the globule, east of cloud C, have been discovered by \citet{Ogu02}.
 The northermost of those coincides with 2MASS J05380259-0134392.

 Based on CO observations \citet{Sug89} found a molecular outflow around
 IRAS~05355$-$0146, which was identified as the driving source. The outflow
 was studied in detail by \citet{Cer92} and was later identified with
 HH~289 by \citet{Mad99} using optical and infrared observations. Such
 images revealed a series of bowshocks in the east-west direction around
 IRAS~05355$-$0146, further confirming that the IRAS source is driving
 the outflow\footnote{Note that in \citet{Mad99} IRAS~05355-0146 is erroneously
 reported as IRAS~05355$-$0416.}. The latter authors also concluded
 that the size of the outflow is slightly larger than 1\,parsec, including
 it in the class of the parsec-scale flows.

 All the above signatures testify that star formation in this region
 may have been triggered by the impact of hot star winds from $\sigma$~Ori.

\subsection{OS98-41}

 This cloud lies about 50~arc-min east of the Trapezium cluster
 (see Figure~\ref{sOri_glob_tails}). Its cometary tail points eastward,
 indicating that the stars of the Trapezium are probably blowing the cloud
 material away. There is a large number of Parenago stars in the field of
 this cloud that have been studied photometrically by \citet{Reb00},
 but no spectroscopic evidence of their youth is available.
 Other emission-line and variable stars, like Haro~4-107 (= Kiso~A-0976~331)
 and V1786~Ori, lie in the field of this cloud. The sources IRAS~05359-0515
 and IRAS~05362-0516 are located in the body of the cloud, which is usually
 identified with the former.
% The cloud is usually associated with the former.

\section{Cloud Distances}
\label{distances}

So far there is not much data on the distances of the Orion outlying clouds.
For some of the clouds the distance to the Orion nebula cluster is normally
assumed\footnote{Note that the distance of 414$\pm$7~pc has been recently reported
by \citet{Men07} for the Orion nebula cluster, which is about 10\% less than
the canonical value.}.
A summary of the available information is given in Table~\ref{tab:dist},
accompanied by a brief presentation of each individual cloud.

%%%%%%%%%%%%%%%%%%%%%%%%%%%%%%%%%%%%%%%%%%%%%%%%%%%%%%%%%%%%%%%%%%%%%%%%%%%%%%%%%%%%%%
\begin{table}[!ht]
\caption{\label{tab:dist} \normalsize Distance of the Orion Outlying Clouds.}
% \label{ic2118clouds}
% \smallskip
 \begin{center}
% {\small
\begin{tabular}{lcl@{\hskip6pt}l}
\tableline
\noalign{\smallskip}
Cloud & Distance  & comments &  refs.  \\
      &  [pc]     &	         &         \\
\noalign{\smallskip}
\tableline
\noalign{\smallskip}
CB~29               & 460-500 & assumed to be the same as L\,1634     &         \\
L~1634              & 460-500 &                                      & 1       \\
L\,1616$^\dagger$    & 450     & based on HD~293815 distance          &         \\
L\,1615$^\dagger$    & 450     & based on HD~293815 distance          &         \\
CB~28               & 450     & assumed to be the same as L\,1615 and L\,1616 &  \\
IC~2118             & 200-230 &     MBM~21                           & 2, 3, 4 \\
LBN~991$^\dagger$   &   ?     &                                      &         \\
LBN~917             & 100-200 &                                      &         \\
LBN~906             & 100-200 &                                      &         \\
L~1642$^\dagger$    & 100-160 & LBN~981, MBM~20                      & 5, 3, 6 \\
\noalign{\smallskip}
\tableline

\multicolumn{4}{l}{\parbox{0.9\textwidth}{\footnotesize
    $^\dagger$ coinciding with an X-ray clump (see Sect.~1).}}\\

\multicolumn{4}{l}{\parbox{0.9\textwidth}{\footnotesize
    References:
    1:~\citet{Fukui89};
    2:~\citet{Franco89};
    3:~\citet{Penp93};
    4:~\citet{Kun01};
    5:~\citet{Penp92};
    6:~\citet{Hearty00}
}}

\end{tabular}
\end{center}
\end{table}
%%%%%%%%%%%%%%%%%%%%%%%%%%%%%%%%%%%%%%%%%%%%%%%%%%%%%%%%%%%%%%%%%%%%%%%%%%%%%%%%%%%%%%

\noindent
{\bf L\,1634, CB~29:}
It has been assumed that L\,1634 is at a distance of 500~pc \citep{Fukui89}.
\citet{Bohi93} argued that L\,1634 is a typical young cloud in Orion and
adopted a distance of 460~pc. However, no direct determinations of the distance
to these clouds are available so far.\\

\noindent
{\bf L\,1616, L\,1615, CB~28:} there are no direct determinations of the distance
to these clouds. However, the distance of the B9V star HD~293815 in L\,1616,
is 450~pc \citep{War78}. Moreover, the isochronal age of 1-2~Myr derived by
\citet{Alc04, Gan08} for the stars in L\,1616 is consistent with the PMS
evolution time of a ZAMS star like HD~293815, when adopting the distance
of 450~pc. This provides further support for the distance of these clouds
to be about 450~pc.
% Moreover, the Herbig Ae/Be star UX~Ori, apparently associated with
% these clouds, is supposed to be at a distance of 450~pc (Natta et al. 2001).
\\

\noindent
{\bf IC~2118:}
% \citet{Kun01} provided several arguments in favor of
\citet{Kun01} argued a distance of 210~pc for the clouds in this region
based on several indications. In particular, the studies by \citet{Franco89}
and \citet{Penp93} of the MBM~21 cloud suggested a distance of that order.
Moreover, the B8Ia-type supergiant {\em Rigel}, which is thought to be the
illuminating source of the reflection nebula, is at a distance of 236~pc.
It is thus reasonable that the distance of the IC~2118 clouds is in the range
between 200 and 230~pc.\\

\noindent
{\bf L\,1642:}
The distance to L\,1642 has been derived using different techniques.
\citet{MagdV86}, from star counts, estimated a  distance of 75--125~pc.
\citet{Penp92} derived a photometric distance between 100 and 120 pc.
Using ROSAT observations, \citet{Kuntz97} reported on the possible detection
of a 0.25\,keV X-ray shadow due to the cloud and suggested that L\,1642 is
within or close to the edge of the Local Bubble, which in that direction is
at about 140~pc, according to \citet{Sfeir99}.
The spectroscopic technique was also used to look for interstellar NaI D lines
signature in bright stars in the direction of the cloud.
\citet{Penp93}, using spectroscopic parallaxes of foreground and background
stars to the cloud, derived an upper limit to the distance of about 110~pc,
while \citet{Hearty00}, using trigonometric parallaxes measured by the
Hipparcos satellite, derived a distance between $112\pm15$pc and $161\pm21$pc.\\

\noindent
{\bf  LBN~917, LBN~906, LBN~991:}
The distance to these clouds is not known. However, at least in the
case of LBN~917 and LBN~906, there are evidences showing that they
might be located inside the Orion-Eridanus Bubble. \citep{Bally91, Brown95}.
It thus seems reasonable that the distance of these two clouds be
in the range between 100 and 200~pc.

\section{Summary}
\label{sum}

As a summary we present in Table~\ref{tab:summary} the numbers of
YSOs and outflows in each region discusses above and a bibliographic guide
to the observational studies performed in different wavelength ranges.

The apparent small number of YSOs found so far in the region of  L\,1634
compared to L\,1616 may result from the fact that the studies conducted in
this region have been mainly devoted to the investigation of the outflow,
rather than to the search of PMS stars. Many more low-mass PMS stars in
L~1634 are hence expected to be identified in the coming years. It is not
surprising that the clouds with most active star formation are closer in
angular distance to the main Orion molecular clouds, in particular L\,1634,
L\,1615 and L\,1616. Star formation in the IC~2118 region is also ongoing,
although not so vigorously as in L\,1616.
IC~2118 may be connected to the star-forming complex in Orion. However,
a true link has not been clarified yet. Given the distance of 210~pc
for the reflection nebula and its associated molecular clouds, the stellar
association in IC~2118 may be located within the Orion--Eridanus Bubble
\citep{Kun04}. The other clouds at higher galactic latitude, like L\,1642,
are probably related to the Orion-Eridanus bubble rather than to the Orion
star forming region. In fact, those clouds are closer to the Sun,
consistently with the idea that they may be associated with the nearest
side of the bubble outskirt. Further investigations of the PMS populations
of these clouds in the coming years will help to shed light on their
relationship with the Orion-Eridanus bubble.

\clearpage

\begin{table}[!ht]
%% S.Rodney 071017 : reduce table column separation to squeeze whitespace
\setlength{\tabcolsep}{0.5\tabcolsep}

\caption{\label{tab:summary} Summary of observations in the Orion outlying clouds and other globules}
\label{coordinates}
% \smallskip
\begin{flushleft}
{\footnotesize
% \begin{tabular}{lrccrccl}
 \begin{tabular}{l@{\hskip3pt}r@{\hskip5pt}c@{\hskip3pt}c@{\hskip8pt}r@{\hskip8pt}c@{\hskip8pt}c@{\hskip8pt}l}
\tableline
\noalign{\smallskip}
                              &   \multicolumn{2}{c}{Number of} & ~ &	 \multicolumn{4}{c}{References~~} \\
%                        &       &          & ~  &          &  &		& \\
 Cloud   & YSOs  & outflows & ~  &   X-rays    &	optical     &	IR  &  radio   \\
\noalign{\smallskip}
\tableline
\noalign{\smallskip}
L~1634, CB~29              & 9      &  2     & ~   & 15, 16     & 1, 5, 10     &  2, 10, 13,     &  3, 22, 23,   \\
                           &        &        & ~   &	        & 12, 16, 43   & 17, 24, 29, 44  &  4, 7, 17     \\
\noalign{\smallskip}
L\,1616 and 1615, CB~28      & 57     & 1      & ~   & 15, 16     & 16, 37, 43  & 16, 25, 30, 37  &  25, 7, 14, 32 \\

\noalign{\smallskip}
IC~2118                    & 10     &        & ~  & 15, 16     & 16, 26, 43    & 21, 30          &  31, 4, 9, 21 \\

\noalign{\smallskip}
LBN~991                    & 1      &        & ~  & 15, 16     & 16	       &   --            &   --	   \\

\noalign{\smallskip}
LBN~917, LBN~906           & 1      &        & ~  &  15	       & 11	       &   30            &   9           \\

\noalign{\smallskip}
L~1642                     & 6      & 1      & ~  &  15	       & 6, 18, 19     &  20, 28, 30     &  3, 4, 27     \\

\noalign{\smallskip}
anonymous X-ray clump      & 4      &        & ~  &  15, 16    & 16	       &     --          &  --           \\

\noalign{\smallskip}
OS98-27                    & 2      &        & ~  &            &  33, 34       &     --          &    --         \\

\noalign{\smallskip}
OS98-35                    & 1      &        & ~  &            &  33           &    --           &    --        \\

\noalign{\smallskip}
OS98-40 = Ori~I-2          & 1      &  1     & ~  &   38       &  35, 40        &    35, 42       &    36, 39, 40, 41 \\

\noalign{\smallskip}
\tableline
\end{tabular}
}
\begin{flushleft}
{\footnotesize
1:~\citet{Cohen80};
2:~\citet{Coh85};
3:~\citet{MagBliMun85};
4:~\citet{Mad86};
5:~\citet{Step86};
6:~\citet{Sand87};
7:~\citet{Clem88};
8:~\citet{ReipHeat90};
9:~\citet{Bally91};
10:~\citet{Sugi91};
11:~\citet{GrHe92};
12:~\citet{Bohi93};
13:~\citet{HoLa95};
14:~\citet{Ram95};
15:~\citet{Sterz95};
16:~\citet{Alc96, Alc00, Alc04};
17:~\citet{Davis97};
18:~\citet{Hearty00};
19:~\citet{LiCh00};
20:~\citet{Vert00};
21:~\citet{Kun01};
22:~\citet{Beltr02};
23:~\citet{DeVr02};
24:~\citet{Nisi02};
25:~\citet{Stank02};
26:~\citet{Kun04};
27:~\citet{Russeil03};
28:~\citet{Leht04};
29:~\citet{OConn04};
30:~\citet{reach98};
31:~\citet{Yon99};
32:~\citet{Park04};
33:~\citet{Yun97};
34:~\citet{Bri05};
35:~\citet{Mad99};
36:~\citet{Sug89};
37:~\citet{Gan08};
38:~\citet{Car98};
39:~\citet{Lar99};
40:~\citet{Cer92};
41:~\citet{Bot96};
42:~\citet{Hod94};
43:~\citet{Lee07};
44:~\citet{Sea08}
}
\end{flushleft}

\end{flushleft}
\end{table}

\vspace{0.5cm}

{\bf Acknowledgements.}
We thank the referee, K. Sugitani, for his careful reading and helpful
comments and suggestions. We are also grateful to Bo Reipurth for his
manifold inputs and comments on an earlier version of the manuscript.
We thank T. Stanke and C. Davis for providing figures from their work.
We are grateful to Steve Rodney for his help with LaTex and the scaling
of the format of some of the figures.
We also thank A. Frasca, L. Spezzi, D. Gandolfi, E. Marilli, L. Testi
and A. Natta for fruitful discussions, as well as for their collaboration
in some of the works mentioned in this chapter.
We acknowledge the use of the color image of IC~2118 by Noel Carboni.
This work was partially financed by the Italian Ministery of University
and Research (MIUR). Financial support from INAF (PRIN-INAF-2005 project
''Stellar clusters: a benchmark for star formation and stellar evolution'')
and from {\em Regione Campania} is also acknowledged. This research has
made use of the SIMBAD database, operated at CDS, Strasbourg, France.

\clearpage

%%% THE BIBLIOGRAPHY
%%%
%%% CONSULT SECTION 3 OF "INSTRUCTIONS FOR AUTHORS" FOR HOW TO USE NATBIB.
%%% AUTHORS ARE ENCOURAGED TO USE EITHER THE "THEBIBLIOGRAPY" ENVIRONMENT
%%% BY UNCOMMENTING (DELETING THE "%" SYMBOL) THE COMMANDS BELOW, OR BY
%%% USING THE BIBTEX ENVIRONMENT. TO FIND OUT WHICH IS APPLICABLE TO YOUR
%%% CONTRIBUTION, CONSULT THE VOLUME EDITORS FOR YOUR PROCEEDINGS.
%%%

%-------------------------------------------------------------------------------------------------------------------------------------------------------------------
\begin{landscape}
\begin{table}[!ht]
\caption{\label{tab:tts} Young stellar objects in the Orion outlying clouds}
\smallskip

{\small
\begin{tabular}{lccrlll}
\tableline
\noalign{\smallskip}
Clouds / Identifier            & $\alpha$ (2000) &   $\delta$ (2000)     &     V    & Sp. Type &  Other Ids.         	  		      &    Refs.      \\
                               &    h~~~m~~~s    &   $^{\circ}$~~~'~~~'' &          &          & 	             	  		      & 	      \\
\noalign{\smallskip}
\tableline

\noalign{\smallskip}
{\bf L 1634 and CB 29:}        &                 &                       &          &          &                     	  		       &	       \\
\noalign{\smallskip}

StHA 36                        & 05:17:46.75     &  $-$03:58:47.0        &   13.00  &	       &	             	  		       &  1, 2, 3       \\
RX\,J0517.9-0708$~^\dagger$    & 05:17:55.00     &  $-$07:08:25.0        &   10.70  &    K2    &                     	  		       &  4, 5, 6       \\
IRAS\,05173-0555               & 05:19:48.90     &  $-$05:52:05.0        &    --    &          & VLA\,3              	  		       &  7, 8, 44      \\
LDN~1634~7                     & 05:19:51.60     &  $-$05:52:06.1        &    --    &          &                     	  		       &  7, 8, 44      \\
StHA\,37                       & 05:20:19.46     &  $-$05:45:55.4        &   13.31  &          & HBC\,83, Kiso\,A-0975\,52, IRAS\,05178-0548  & 1, 9, 10, 11, 3, 43 \\
StHA\,38$~^\dagger$            & 05:20:25.75     &  $-$05:47:06.4        &   14.50  &          & V\,534\,Ori, Kiso\,A-0975\,54                & 1, 12, 10, 2, 3, 43 \\
StHA 39                        & 05:20:31.43     &  $-$05:48:24.6        &   13.50  &          &      RX\,J0520.5-0548                        & 1, 6, 2, 3, 43   \\
RX\,J0520.9-0452               & 05:20:56.00     &  $-$04:52:43.0        &    9.53  &    F7    &                                              & 4, 5, 6	        \\
Kiso\,A-0975\,69$~^\dagger$    & 05:23:03.20     &  $-$04:40:37.0        &   14.68  &    K6    &     RX\,J0523.1-0440	                      & 10, 6, 4, 13    \\

\noalign{\smallskip}
{\em Candidates:}              &                 &	 	         &	    &          &		          		       &	       \\
\noalign{\smallskip}

Kiso\,A-0975\,40               & 05:17:52.25     &  $-$03:26:01.0        &   18.00  &	       &                          		       & 10	      \\
Kiso\,A-0975\,43               & 05:18:16.85     &  $-$05:37:30.0        &   15.70  &	       &	                  		       & 10	      \\
Kiso\,A-0975\,45               & 05:19:13.56     &  $-$03:24:12.6        &   12.50  &	       &	                  		       & 10	      \\

VLA\,1                         & 05:19:44.02     &  $-$05:54:13.2        &     --   &          &                          		       & 14	      \\
VLA\,2                         & 05:19:45.36     &  $-$05:52:36.6        &     --   &          &                          		       & 14	      \\
VLA\,4                         & 05:19:54.77     &  $-$05:55:39.4        &     --   &          &                          		       & 14	      \\
VLA\,5                         & 05:19:58.92     &  $-$05:53:49.9        &     --   &          &                          		       & 14	      \\

Kiso\,A-0975\,86               & 05:25:39.79     &  $-$04:11:02.0        &   15.50  &  	       & 	                  		       & 10	      \\

\noalign{\smallskip}
\tableline
\noalign{\smallskip}

{\bf L\,1616}, {\bf L\,1615}, {\bf CB 28:} &    &                       &          &          &                           		       &	     \\
\noalign{\smallskip}
 1RXS\,J045912.4$-$033711      & 04:59:14.59    &  $-$03:37:06.3         &  11.73   &    G8    &                            		      & 17,15       \\
 1RXS\,J050416.9$-$021426      & 05:04:15.93    &  $-$02:14:50.5         &  12.96   &    K3    &                            		      & 17,15	    \\
 UX\,Ori$~^\dagger$            & 05:04:29.99    &  $-$03:47:14.3         &   9.61   &   A3e    & HBC\,430, IRAS\,05020-0351, HD\,293782       & 11, 16      \\
 TTS\,050513.5$-$034248        & 05:05:13.47    &  $-$03:42:47.8         &    --    &    M5.5    &                                	      & 17          \\
 TTS\,050538.9$-$032626        & 05:05:38.85    &  $-$03:26:26.4         &  17.30   &    M3.5    &                                	      & 17          \\
 RX\,J0506.6$-$0337 	       & 05:06:34.95    &  $-$03:37:15.9         &  12.23   &    G7    &                                	      & 17,15	    \\
\noalign{\smallskip}
\tableline

\end{tabular}
}
\end{table}
\end{landscape}

%-------------------------------------------------------------------------------------------------------------------------------------------------------------------

\begin{landscape}
\begin{table}[!ht]
{\small ~~~~~~~~Table~\ref{tab:tts}. Continued}
\smallskip

{\small
\begin{tabular}{lccrlll}
\tableline
\noalign{\smallskip}
Clouds / Identifier           & $\alpha$ (2000) &   $\delta$ (2000)      &     V    &  Spectral &        Other               &    Refs.  \\
                              &    h~~~m~~~s   &   $^{\circ}$~~~'~~~''   &          &  Type     & 	 identification      &          \\
\noalign{\smallskip}
\tableline
\noalign{\smallskip}
 TTS\,050644.4$-$032913        & 05:06:44.42    &  $-$03:29:12.8         &  17.69   &    M4.5    &                             & 17                             \\
 TTS\,050646.1$-$031922	       & 05:06:46.05    &  $-$03:19:22.4         &  17.34   &    K4    &                             & 17,15                          \\
 RX\,J0506.8$-$0318 	       & 05:06:46.64    &  $-$03:18:05.6         &  14.86   &    K8.5    &                             & 17,15        	              \\
 TTS\,050647.5$-$031910        & 05:06:47.45    &  $-$03:19:09.7         &  20.51   &   M5.5   &                             & 17                             \\
 RX\,J0506.8$-$0327 	       & 05:06:48.32    &  $-$03:27:38.2         &  15.96   &    M3.5    &                             & 17,15        	              \\
 RX\,J0506.8$-$0305            & 05:06:48.98    &  $-$03:05:42.9         &  17.33   &    M4.5    &                             & 17  		              \\
 TTS\,050649.8$-$031933        & 05:06:49.77    &  $-$03:19:33.1         &  17.63   &    M3.5    &		             & 17  		              \\
 TTS\,050649.8$-$032104        & 05:06:49.78    &  $-$03:21:03.6         &  18.59   &    M1    &		             & 17  		              \\
 TTS\,050650.5$-$032014        & 05:06:50.50    &  $-$03:20:14.3         &  21.11   &   M6.5   &		             & 17  		              \\
 TTS\,050650.7$-$032008        & 05:06:50.74    &  $-$03:20:08.0         &  19.54   &   M4.5   &		             & 17  		              \\
 RX\,J0506.9$-$0319~NW         & 05:06:50.83    &  $-$03:19:35.2         &  16.73   &    M3    &                             & 17,15         	              \\
 RX\,J0506.9$-$0319~SE         & 05:06:50.99    &  $-$03:19:38.0         &  14.82   &    K5    &     L\,1616\,MIR5           & 17,15,18      	              \\
 HD\,293815		       & 05:06:51.05    &  $-$03:19:59.9         &  10.08   &    B9    &                             & 17,15,19                       \\
 RX\,J0506.9$-$0320W	       & 05:06:52.86    &  $-$03:20:53.2	 &  15.33   &	 K8.5    &	L\,1616\,MIR2	     & 17,15,18		              \\
 RX\,J0506.9$-$0320E	       & 05:06:53.32    &  $-$03:20:52.6	 &  15.74   &	 K1    &	L\,1616\,MIR1	     & 17,15,18		              \\
 TTS\,050654.5$-$032046        & 05:06:54.53    &  $-$03:20:46.0         &   --     &   M4   &                             & 17                             \\
 LkH$\alpha$~333$~^\dagger$    & 05:06:54.65    &  $-$03:20:04.8          &  14.19   &    K4    & HBC 82, RX\,J0506.9-0320, Kiso\,A-0974-14  & 17,15, 20, 11, 43 \\
 L\,1616\,MIR4	               & 05:06:54.93    &  $-$03:21:12.7         &  18.99   &    K1    &                             & 17, 13, 12, 18	              \\
 Kiso\,A-0974-15	       & 05:06:55.52    &  $-$03:21:13.2         &  12.84   &    B3    & NSV\,1832, L1616\,MIR3, IRAS\,05044-0325  & 17, 15,18, 21, 20, 22, 23   \\
 RX\,J0507.0$-$0318 	       & 05:06:56.94    &  $-$03:18:35.5	 &  15.21   &	 M0    &			     & 17,15	                     \\
 TTS\,050657.0$-$031640        & 05:06:56.97    &  $-$03:16:40.4         &  17.69   &   M4.5   &	                     & 17                            \\
 TTS\,050704.7$-$030241        & 05:07:04.71    &  $-$03:02:41.0         &  20.04   &    M6    &  	                     & 17                            \\
 TTS\,050705.3$-$030006        & 05:07:05.32    &  $-$03:00:06.2         &  15.50   &   M0   &	                     & 17                            \\
 RX\,J0507.1$-$0321 	       & 05:07:06.10    &  $-$03:21:28.2         &  16.13   &	M1   &	Kiso\,A-0974-16      & 17,15,18, 20, 43	             \\
 TTS\,050706.2$-$031703        & 05:07:06.22    &  $-$03:17:02.9         &  19.99   &   M6   &                             & 17                            \\
 RX\,J0507.2$-$0323 	       & 05:07:10.95    &  $-$03:23:53.4         &  13.95   &    K4    &	   Kiso\,A-0974-18   & 17,15,18, 20	             \\
\noalign{\smallskip}
\tableline

\end{tabular}
}
\end{table}
\end{landscape}

\begin{landscape}
\begin{table}[!ht]
{\small ~~~~~~~~Table~\ref{tab:tts}. Continued}
\smallskip

{\small
\begin{tabular}{lccrlll}
\tableline
\noalign{\smallskip}
Clouds / Identifier           & $\alpha$ (2000) &   $\delta$ (2000)     &     V      &  Spectral &        Other               &    Refs.                   \\
                              &    h~~~m~~~s    &   $^{\circ}$~~~'~~~''  &           &  Type     & 	 identification       &                            \\
\noalign{\smallskip}
\tableline
\noalign{\smallskip}
 TTS\,050713.5$-$031722       & 05:07:13.52     &  $-$03:17:22.1         &  17.57    &    K8.5    &                             & 17                        \\
 RX\,J0507.3$-$0326           & 05:07:14.99     &  $-$03:26:47.3         &  14.31    &    M0    &                             & 17                        \\
 TTS\,050717.9$-$032433	      & 05:07:17.85     &  $-$03:24:33.1         &  16.67    &    M2.5    &                             & 17,15  	                  \\
 RX\,J0507.4$-$0320 	      & 05:07:22.28     &  $-$03:20:18.5         &  16.71    &    M4    &                             & 17,15  	                  \\
 RX\,J0507.4$-$0317$^a$       & 05:07:25.93     &  $-$03:17:12.3         &  17.19    &	  M3    & 	                      & 17  			  \\
 TTS\,050729.8$-$031705       & 05:07:29.80     &  $-$03:17:05.1         &  20.83    &   M6.5   & 	                      & 17  			  \\
 TTS\,050730.9$-$031846       & 05:07:30.85     &  $-$03:18:45.6         &  22.12    &   M5.5   &		              & 17  			  \\
 TTS\,050733.6$-$032517       & 05:07:33.58     &  $-$03:25:16.7         &  19.62    &    M5.5    &  	                      & 17  			  \\
 TTS\,050734.8$-$031521       & 05:07:34.83     &  $-$03:15:20.7         &  19.05    &   M5   &  	                      & 17  			  \\
 RX\,J0507.6$-$0318$^a$       & 05:07:37.67     &  $-$03:18:15.6         &  15.16    &    K7    &  	                      & 17  			  \\
 TTS\,050741.0$-$032253       & 05:07:41.00     &  $-$03:22:53.0         &  17.55    &   M4   &  	                      & 17  			  \\
 TTS\,050741.4$-$031507       & 05:07:41.35     &  $-$03:15:06.7         &  17.92    &    M4.5    &  	                      & 17  			  \\
 TTS\,050752.0$-$032003       & 05:07:51.95     &  $-$03:20:02.8         &  19.55    &    M5.5    &  	                      & 17  			  \\
 TTS\,050801.4$-$032255       & 05:08:01.43     &  $-$03:22:54.5         &    --     &    M0.5    &                             & 17,15  	                  \\
 TTS\,050801.9$-$031732       & 05:08:01.94     &  $-$03:17:31.6         &    --     &    M1    &                             & 17,15  	                  \\
 TTS\,050804.0$-$034052       & 05:08:04.00     &  $-$03:40:51.7         &    --     &    M2.5    &                             & 17,15  	                  \\
 TTS\,050836.6$-$030341       & 05:08:36.55     &  $-$03:03:41.4         &    --     &    M1.5    & Kiso\,A-0974-21             & 17,15         	          \\
 TTS\,050845.1$-$031653       & 05:08:45.10     &  $-$03:16:52.5         &    --     &    M3.5    &                             & 17,15         	          \\
 RX\,J0509.0$-$0315$~^\dagger$& 05:09:00.66     &  $-$03:15:06.6         &  11.39    &    G8    & 1RXS\,J050859.6$-$0315$-$03 & 17,15         	         \\
 RX\,J0510.1$-$0427$~^\dagger$& 05:10:04.60     &  $-$04:28:03.7         &  11.73    &    K4    &   1RX\,J051004.9$-$042757   & 17,15         	         \\
 1RXS~J051011.5$-$025355      & 05:10:10.86     &  $-$02:54:04.9         &  12.42    &    K0    & V1011~Ori, IRAS\,07076-0257 & 17,15, 24, 12             \\
 RX\,J0510.3$-$0330$~^\dagger$& 05:10:14.78     &  $-$03:30:07.4         &  11.74    &    G8    &   1RXS\,J051015.7-033001    & 17,15, 4, 6	         \\
 1RXS\,J051043.2$-$031627     & 05:10:40.50     &  $-$03:16:41.6         &  11.38    &    G2    &                             & 17,15  	                  \\
 RX\,J0511.7$-$0348$~^\dagger$& 05:11:38.93     &  $-$03:48:47.1         &  12.02    &    K1    &                             & 17,15, 4, 6	         \\
 RX\,J0512.3$-$0255$~^\dagger$& 05:12:20.53     &  $-$02:55:52.3         &  12.61    &    K2    &         V531~Ori            & 17,15, 4, 12, 24, 6, 43\\
 L\,1616\,MMS1\,A	      & 05:06:44.40     &  $-$03:21:34.0         &  --       &    --    &		              & 18	                 \\
 L\,1616\,MMS1\,B  	      & 05:06:43.70     &  $-$03:21:28.0         &  --       &    --    &			      & 18	                 \\

\noalign{\smallskip}
\tableline

\end{tabular}
}
\end{table}
\end{landscape}

%-------------------------------------------------------------------------------------------------------------------------------------------------------------------

\begin{landscape}
\begin{table}[!ht]
{\small ~~~~~~~~Table~\ref{tab:tts}. Continued}
\smallskip

{\small
\begin{tabular}{lccrlll}
\tableline
\noalign{\smallskip}
Clouds / Identifier                      & $\alpha$ (2000) &   $\delta$ (2000)    &   V    &  Spectral &        Other            &    Refs.         \\
                                         &    h~~~m~~~s   &   $^{\circ}$~~~'~~~'' &         &  Type    & 	 identification  &                  \\
\noalign{\smallskip}
\tableline
\noalign{\smallskip}
 L\,1616\,MMS1\,C	                  & 05:06:43.40 &  $-$03:21:38.0    &  --      &    --      &		                  & 18              \\
 L\,1616\,MMS1\,D	                  & 05:06:42.90 &  $-$03:21:31.0    &  --      &    --      &				  & 18              \\

\noalign{\smallskip}

\tableline
\noalign{\smallskip}
{\bf IC2118:}                              &            &	 	   &	       &          &                              &                  \\
\noalign{\smallskip}

 RX\,J0500.4-1054                         & 05:00:25.83 & $-$10:54:22.3    &   13.00  &    K7    &                              & 6, 4, 5           \\
 IRAS\,04591-0856                         & 05:01:30.20 & $-$08:52:14.0    &    --    &          &  G\,13, HHL\,17              & 25, 26            \\
 2MASS\,J05020630$-$0850467$~^\dagger$    & 05:02:06.31 & $-$08:50:46.6    &    --    &    M2IV  &                              & 26, 27            \\
 RX\,J0502.4-0744$~^\dagger$              & 05:02:20.84 & $-$07:44:09.9    &   11.21  &    G6    &                              & 6, 4, 5           \\
 RX\,J0503.8-1130$~^\dagger$              & 05:03:49.55 & $-$11:31:01.0    &    --    &    K1    &                              & 6, 4, 5           \\
 2MASS\,J05065349$-$0617123$~^\dagger$    & 05:06:53.51 & $-$06:17:12.5    &    --    &    K7IV  &  IRAS\,F\,05044-0621         & 26, 27            \\
 2MASS\,J05071157$-$0615098$~^\dagger$    & 05:07:11.57 & $-$06:15:10.0    &    --    &    M2IV  &  IRAS\,F\,05047-0618         & 26, 27            \\
 2MASS\,J05073016$-$0610158$~^\dagger$    & 05:07:30.18 & $-$06:10:15.8    &    --    &    K6IV  &  IRAS\,05050-0614, Kiso\,A-0974\,19 & 26, 27, 10, 43 \\
 2MASS\,J05073060$-$0610597$~^\dagger$    & 05:07:30.62 & $-$06:10:59.7    &    --    &    K7IV  &                              & 26, 27, 43        \\
 RX\,J0507.8-0931$~^\dagger$              & 05:07:48.33 & $-$09:31:43.2    &    --    &    K2    &                              & 6, 4, 5           \\

\noalign{\smallskip}

\noalign{\smallskip}
{\em Candidates in IC\,2118:}              &             &	 	    &	       &           &		                  &                  \\
\noalign{\smallskip}

2MASS J05060574$-$0646151       	   & 05:06:05.75 &  $-$06:46:15.2   &	 --    &    G8~:  &				  & 26  	    \\
2MASS J05094864$-$0906065       	   & 05:09:48.65 &  $-$09:06:06.6   &	 --    &    G8    &				  & 26  	    \\
2MASS J05112460$-$0818320       	   & 05:11:24.60 &  $-$08:18:32.1   &	 --    &    M0    &				  & 26  	    \\

\noalign{\smallskip}
\tableline
\noalign{\smallskip}
{\bf L\,1642:}                            &               &	 	     &	         &          &                              &                 \\
\noalign{\smallskip}

BD-15 808$~^\dagger$            	  & 04:32:43.51   & $-$15:20:11.3   &	10.39  &   G4V    & GSC\,05891-00069, 1RXS\,J043243.2-152003 &  28	      \\
L 1642-2$~^\dagger$             	  & 04:34:49.73   & $-$14:13:08.1   &	17.00  &    --    & HBC\,410, IRAS\,04325-1419  	    &  11, 29, 30, 31 \\
L\,1642-2B                      	  & 04:34:49.98   & $-$14:13:12.8   &	 --    &    --    & 2MASS\,J04344997-1413128		    &	    29        \\
L\,1642-1$~^\dagger$            	  & 04:35:02.29   & $-$14:13:40.8   &	13.70  &   K7IV   & EW Eri, HBC\,413, IRAS\,04327-1419      &  11, 29, 30     \\
2MASS J04351455-1414468         	  & 04:35:14.55   & $-$14:14:46.9   &	 --    &    --    & 2MASSI\,J0435145-141446	            &  32	      \\
IRAS 04336-1412                 	  & 04:35:55.30   & $-$14:05:58.0   &	 --    &    --    &  L\,1642-3  			    & 29, 33	      \\
IRAS 04347-1415                 	  & 04:37:03.40   & $-$14:09:01.0   &	 --    &    --    &  L\,1642-4  			    & 29, 33	      \\

\noalign{\smallskip}
\tableline

\end{tabular}
}

\end{table}
\end{landscape}

%-------------------------------------------------------------------------------------------------------------------------------------------------------------------
\begin{landscape}
\begin{table}[!ht]
{\small ~~~~~~~~Table~\ref{tab:tts}. Continued}
\smallskip

{\small
\begin{tabular}{lccrlll}
\tableline
\noalign{\smallskip}
Clouds / Identifier                      & $\alpha$ (2000) &   $\delta$ (2000)    &     V   &  Spectral   &        Other      &    Refs.                   \\
                                         &    h~~~m~~~s   &   $^{\circ}$~~~'~~~'' &          &  Type       & 	 identification  &                         \\
\noalign{\smallskip}
\tableline
\noalign{\smallskip}
{\em Candidates:}                        &               &	 	          &	     &         &                         &	                      \\
\noalign{\smallskip}

IRAS 04342-1444                          & 04:36:33.88   & $-$14:38:58.0          &    9.03  &	K5     & BD-14 929, GSC\,05327-00570, HIP\,21463 & 33        \\
IRAS 04349-1436                          & 04:37:13.20   & $-$14:30:34.0	  &	     &	       &					 & 33        \\

\tableline
\noalign{\smallskip}
{\bf LBN\,991:}                         &                &	 	           &	      &          &                       &                            \\
\noalign{\smallskip}

RX\,J0513.4-1244$~^\dagger$             &  05:13:22.02   & $-$12:44:52.4          &  10.70   &   G4     &  GSC 05338-00490        & 6, 4                     \\

\tableline
\noalign{\smallskip}
{\bf LBN\,917}, {\bf LBN\,906:}         &                &	 	           &	      &           &                         &                        \\
\noalign{\smallskip}

IRAS\,04451-0539                        &  04:47:34.10   & $-$05:34:14.0           &  14.76   &           &   PDS\,13               &    34                  \\

\tableline
\noalign{\smallskip}
{\bf Anonymous X-ray clump at}         & $\alpha \approx 5^h21^m$ & $\delta \approx -9^\circ$ {\bf :} &	      &      &      &                               \\
\noalign{\smallskip}

RX\,J0515.6-0930$~^\dagger$            &   05:15:36.30    & $-$09:30:51.8 &   9.79  &	 G5	&		    				 &  6, 4       \\
RX\,J0522.1-0844                       &   05:22:03.40    & $-$08:44:19.0 &  12.09  &	 K0	&		    				 &  6, 4       \\
RX\,J0523.0-0850$~^\dagger$            &   05:22:57.00    & $-$08:50:11.5 &   --    &  K7-M0	&		    				 &  6, 4       \\
IRAS\,05222-0844                       &   05:24:37.00    & $-$08:42:00.0 &   9.88  &	 G3	&   PDS\,111, BD-08\,1115, 1RXS J052437.4-084200  &  35        \\

\tableline
\noalign{\smallskip}
{\bf OS98-27, 35, 40, 41: }  Globules    &  around  $\sigma~Ori$ &	 	     &	         &                 &                               &        \\
\noalign{\smallskip}

YSO CB031YC1                           &  05:30:45.00	  & $-$00:38:15.00 & 14.10  &	 M1	&		   			   &  36          \\
CVSO 121                               &  05:33:39.82	  & $-$00:38:54.20 & 14.30  &	 K3	&		   			   &  37          \\
YSO CB032YC1                           &  05:36:30.64	  & $-$00:18:43.60 & 14.0:  &	 --	&		   			   &  36          \\
IRAS~05355$-$0146                      &  05:38:04.90	  & $-$01:45:09.00 & --     &   --	&		   			   & 38           \\

\noalign{\smallskip}
{\em Candidates:}                     &                  &	 	  &	    &           &	                  	           &              \\
\noalign{\smallskip}

Kiso~A-0904~10                        &  05:33:00.60	 & $-$00:36:21.00 & 13.20  &	--     &		  			   & 40  	 \\
IRAS~05337-0019                       &  05:36:20.90	 & $-$00:17:16.00 & 10.37  &	--     & TYC 4766-2306-1  			   & 42          \\
OriI-2N-1  		              &  05:37:51.10	 & $-$01:36:17.00 &  --    &	--     &		  			   & 39 	 \\
OriI-2N-2  		              &  05:37:59.10	 & $-$01:36:56.00 &  --    &	--     &		  			   & 39 	 \\
OriI-2N-3  		              &  05:38:02.40	 & $-$01:36:35.00 &  --    &	--     &		  			   & 39 	 \\
OriI-2N-4  		              &  05:38:02.60	 & $-$01:34:40.00 &  --    &	--     & 2MASS J05380259-0134392                   & 39 	 \\
Kiso~A-0976~331                       &  05:38:25.00	 & $-$05:22:06.00 & 16.00  &   --      & Haro~4-107, PACH~468                      & 40, 41 	 \\

% \noalign{\smallskip}
\tableline

\end{tabular}
}
% \end{center}
\end{table}

\vspace{-0.5cm}
{\footnotesize
\noindent
$^\dagger$: proper motion  available in \citet{Ducour05}\\
References:
{\bf 1.}~\citet{Mah03}; {\bf 2.}~\citet{Dow88}; {\bf 3.}~\citet{Step86}; {\bf 4.}~\citet{Alc00};
{\bf 5.}~\citet{Alc98}; {\bf 6.}~\citet{Alc96}; {\bf 7.}~\citet{HoLa95}; {\bf 8.}~\citet{Davis97};
{\bf 9.}~\citet{Wea92}; {\bf 10.}~\citet{Wir91}; {\bf 11.}~\citet{HeBe88}; {\bf 12.}~\citet{Kin00};
{\bf 13.}~\citet{Cha00}; {\bf 14.}~\citet{Beltr02}; {\bf 15.}~\citet{Alc04}; {\bf 16.}~\citet{Natta99, Natta00};
{\bf 17.}~\citet{Gan08}; {\bf 18.}~\citet{Stank02}; {\bf 19.}~\citet{Nes95}; {\bf 20.}~\citet{Nak95};
{\bf 21.}~\citet{Yon99}; {\bf 22.}~\citet{Coh89}; {\bf 23.}~\citet{Oln86};  {\bf 24.}~\citet{Cie97};
{\bf 25.}~\citet{Gyu82, Gyu87}; {\bf 26.}~\citet{Kun04}; {\bf 27.}~\citet{Kun03}; {\bf 28.}~\citet{LiCh00};
{\bf 29.}~\citet{Sand87}; {\bf 30.}~\citet{OB95};  {\bf 31.}~\citet{Magn95}; {\bf 32.}~\citet{Cruz03};
{\bf 33.}~\citet{Leht04}; {\bf 34.}~\citet{GrHe92}; {\bf 35.}~\citet{GrHe02};
{\bf 36.}~\citet{Yun97};
{\bf 37.}~\citet{Bri05};
{\bf 38.}~\citet{Mad99};
{\bf 39.}~\citet{Ogu02};
{\bf 40.}~\citet{Wir89};
{\bf 41.}~\citet{Har49};
{\bf 42.}~\citet{Mag03};
{\bf 43.}~\citet{Lee07}
{\bf 44.}~\citet{Sea08}
 }

\end{landscape}
%-------------------------------------------------------------------------------------------------------------------------------------------------------------------

\end{document}